\documentclass[a4paper,11pt]{article}
\usepackage[latin1]{inputenc}
\usepackage{amsmath,amsthm,amsfonts,amssymb}
\usepackage{enumerate}
\usepackage{graphicx}
\usepackage{float}

\usepackage{etex}
\usepackage{tikz}
\usetikzlibrary{decorations.markings}
\tikzset
{
   middlearrow/.style=
   {
   decoration={markings,mark= at position 0.5 with {\arrow{#1}} ,},
   postaction={decorate}
   }
}
\usepackage{nccmath}
\usepackage{mathtools}

\usepackage{authblk}

\usepackage[top=1.3in, bottom=1.3in, left=1.35in, right=1.35in]{geometry}
\usepackage{relsize}

\usepackage{array}

\theoremstyle{plain}

\theoremstyle{definition}

\numberwithin{equation}{section}

\title{Modeling trait-dependent evolution\\ on a random species tree}
\author[1,$\ast$]{Daniah Tahir}
\author[2,3]{Sylvain Gl{\'e}min}
\author[2]{Martin Lascoux}
\author[1]{Ingemar Kaj}
\affil[1]{Department of Mathematics, Uppsala University, 751 06 Uppsala, Sweden}
\affil[2]{Department of Plant Ecology and Evolution, Uppsala University, 752 36 Uppsala, Sweden}
\affil[3]{UMR 5554 ISEM, Universit{\'e} de Montpellier, 34095 Montpellier Cedex 5, France}
\affil[$\ast$]{Correspondence to be sent to: Department of Mathematics, Uppsala University, 751 06 Uppsala, Sweden. 
E-mail:~daniah.tahir@math.uu.se}
\date{}                    
\begin{document}

\maketitle

\begin{abstract} 
Understanding the evolution of binary traits, 
which affects the birth and survival of species 
and also the rate of molecular evolution, remains challenging. 
A typical example is the evolution of mating systems in plant species. 
In this work, we present a probabilistic modeling framework for 
binary trait, random species trees, in which the number of species 
and their traits are represented by a two-type, continuous time 
Markov branching process. We develop our model by considering 
the impact of mating systems on $dN/dS$, the ratio of nonsynonymous 
to synonymous substitutions. A methodology is introduced which 
enables us to match model parameters with parameter estimates 
from phylogenetic tree data. The properties obtained from the model are
applied to outcrossing and selfing species trees in the Geraniaceae and 
Solanaceae family. This allows us to investigate not only the branching tree 
rates, but also the mutation rates and the intensity of selection. 
[mathematical modeling; branching processes; phylogenetic trees; mating systems; $dN/dS$.]
\end{abstract}

\section{Introduction}

One of the main obstacles in modern evolutionary biology entails recognizing 
how different traits evolve on a phylogenetic tree, and how they can affect the 
evolution of other species characteristics. A popular example is the impact of life 
history or ecological traits on molecular evolutionary rates (\cite{nikolaev_etal};
\cite{smith_donoghue}; \cite{figuet_etal}). Another subtle
hurdle is the difficulty to estimate precisely when a species moved
from one character state to another, and how irreversible the
change was. These questions have turned out to 
be particularly difficult to answer; in part, it is simply
because what we can observe today, is only what survived, thereby
providing a biased view of the evolutionary process. A classical 
approach involves mapping traits on a phylogenetic
tree, and then matching them with the evolutionary rates that have been 
measured on the corresponding branches. In such approaches, it is 
usually assumed that the phylogenetic tree is fixed, and is independent of
the evolving traits. Yet, it is well known that some traits can affect the
diversification process, and hence, the tree shape, which can bias
the reconstruction of ancestral states as well as state transitions along the
tree \cite{goldberg_igic2008}. Another strong, and often
implicit, assumption is that the change in states only occurs at
branching points. In reality, transition may occur
along the branches, and this results in diluting the relationship
between traits and molecular rates.

A minimal model, addressing the differences in speciation rates
between groups, will therefore need to consider, simultaneously, trait
evolution and species diversification.  The binary-state speciation
and extinction (BiSSE) model was precisely developed for this purpose, 
and it also helped disentangle the two aspects \cite{maddison_etal}. It was later
extended to the ClaSSE (cladogenetic-state speciation and extinction)
model. Generally, binary-state phylogenetic models are
structured in classes satisfying one or several of \cite{goldberg_igic}:

\noindent i) independent speciation and extinction for the two states;

\noindent ii) anagenetic change, i.e. instantaneous state change occurring along lineages;

\noindent iii) cladogenetic change, i.e. change of state occurring during speciation.

Figure \ref{biss} gives an illustration of 
cladogenetic and anagenetic state changes, in panel a) 
and panel b), respectively. A BiSSE model satisfies i) and ii), whereas a ClaSSE model shows all
properties i), ii) and iii). These models have been extensively used 
to test for the effect of traits on species diversification. However, to our 
knowledge, the assessment of the effect of traits on molecular evolutionary 
rates has neglected to take into account the possible effect of traits on species diversification.

A typical example, where both approaches have been conducted
independently, is the evolution of selfing from 
outcrossing, in the context of the so-called `dead-end' 
hypothesis \cite{stebbins}. The popular claim that the
transition towards selfing is an evolutionary dead-end
\cite{busch_igic}, relies on two hypotheses: the transition
from outcrossing to selfing is irreversible, 
and, selfing species go extinct more often than the
outcrossing ones. Both outcrossers and selfers reproduce through
the processes of meiosis and fertilization.  In the outcrossing
species, random fertilization of gametes from distinct individuals
occurs, whereas in selfers, gametes from the same hermaphrodite
individual fuse together to produce new individuals. 
In the short term, selfing offers reproductive assurance and transmission advantage 
over outcrossers, since selfers can contribute
to outcross pollen while fertilizing their own ovules at the same
time \cite{charlesworth}. On the other hand, outcrossing limits a plant's
ability to reproduce in case of rarity of mates. Therefore, the transition from
outcrossing to selfing is thought to be one of the most frequent
evolutionary changes in angiosperms. In the long run, however, selfing
species suffer from negative genetic consequences: selfing reduces
effective population sizes and recombination, which globally lessens
the efficacy of selection. Selfing species are thus, more prone to
the accumulation of deleterious mutations and less able to adapt to
changing environments, which eventually drives them towards
extinction \cite{wright_etal}. In agreement with this prediction and using the
BiSSE model, the net diversification rates (difference between
speciation and extinction rates) are found to be higher in outcrossing
than in selfing Solanaceae species \cite{goldberg_etal}. 
Generally, it is proposed that the net diversification rate in selfers is negative
\cite{busch_igic}.

In parallel, the negative genetic effects of selfing, which can 
explain higher extinction rates, were tested by comparing the molecular 
evolutionary rates between selfing and outcrossing lineages. The 
efficacy of selection can be assessed through the ratio of 
nonsynonymous to synonymous substitutions, $dN/dS$; a higher ratio 
corresponding to less efficient selection. Even though reduced selection efficacy in 
selfers was often detected at the within-species level, phylogenetic 
analyses usually failed at detecting any effect of mating system on 
$dN/dS$ \cite{glemin_muyle}. The recent origin 
of selfing and the misspecification of shift in mating systems, are 
generally recognized as the most likely explanations. Yet, the BiSSE or 
ClaSSE processes underlying the observed trees, have not been 
incorporated so far, in such $dN/dS$ analyses.

In this work, building on the BiSSE and ClaSSE models, we
present a probabilistic modeling framework for binary trait, random
species trees, with trait-dependent mutation rates. Similar models have
been used in statistical inference, but, to our knowledge, their
detailed mathematical properties have not been studied yet. 
As in the BiSSE and ClaSSE models, the edges of our tree model are
grouped into two categories based on a generic trait.  Over time, the
trait has influenced and shaped the ancestral family tree of the
extant species and led eventually to the observable mixture of traits
associated with the present tree branches.  The model specifies
expected values of various random functionals of the tree in terms of
basic Markov chain parameters.  Not only the number of species of each
trait, but also the number of trait-clusters in the tree, as well as
the total branch lengths in the ancestral tree, with existing ancestors
at the time of observation, are provided.

While the tree model is constructed in forward time, it is the traits
and mutations in the ancestral tree seen backwards from the current
set of species, that determine the current set of states. In the model,
this shift of view corresponds to studying the reduced branching tree,
obtained from the original species tree by the removal of extinct
species.  The rate of fixation, as the species accumulate mutations, may
depend on the trait value, due, for example, to a varying degree of
selection associated with the traits. Also, considering mutations to
be deleterious, the rate of extinction of a species could be
a function of the rate of fixation of mutations, in the corresponding
trait. The model allows us to study the accumulated number of
mutations in relation to the observed distribution of traits.

The analysis helps understand the interplay between the evolution of
the trait on one hand, and the mutation activity on the tree branches,
on the other. To test our approach with regards to empirical data, we
discuss a methodology for matching the model parameters with parameter
estimates, using data of reconstructed phylogenetic species trees and
draw useful inferences on the trait dynamics, specifically with
regards to $dN/dS$. The approach relies on finding the expected size
of the reduced tree \cite{nee_etal}. These methods are applied,
in particular, to plant families composed of a mixture of outcrossing (assigned trait-$0$)
and selfing (assigned trait-$1$) species, where $dN/dS$ is presumed to depend on the trait.

\section{Modeling the Species Tree}

\subsection*{Parametrized Branching Tree}

\begin{figure}[!t]
\centerline{
\begin{tikzpicture}[scale=0.8]
\node at (-0.7,3.3) {a)};
\draw [thick,<-]  node[anchor=north]{$0$};
\draw[blue,thick,middlearrow={<}] (0,0) -- (1.1,2.5) ;
\draw[blue,thick,middlearrow={>}] (1.1,2.5) -- (2.2,0)node[black,anchor=north]{$0$};
\draw[blue,thick] (1.1,2.5) -- (1.1,2.8) node[black,anchor=south]{$0$};
\draw[thick] (3,0)  node[anchor=north]{$1$};
\draw[red,thick,middlearrow={<}] (3,0) -- (4.1,2.5) ;
\draw[red,thick] (4.1,2.5) -- (4.1,2.8) node[black,anchor=south]{$1$};
\draw[red,thick,middlearrow={>}] (4.1,2.5) -- (5.2,0)node[black,anchor=north]{$1$};
\draw[thick,black] (6,0)  node[anchor=north]{$0$};
\draw[blue,thick,middlearrow={<}] (6,0) -- (7.1,2.5) ;
\draw[blue,thick] (7.1,2.5) -- (7.1,2.8) node[black,anchor=south]{$0$};
\draw[red,thick,middlearrow={>}] (7.1,2.5) -- (8.2,0)node[black,anchor=north]{$1$};
\draw[black,thick] (9,0)  node[anchor=north]{$0$};
\draw[blue,dashed,thick,middlearrow={<}] (9,0) -- (10.1,2.5); 
\draw[red,thick] (10.1,2.5) -- (10.1,2.8) node[black,anchor=south]{$1$};
\draw[red,dashed,thick,middlearrow={>}] (10.1,2.5) -- (11.2,0) node[black,anchor=north]{$1$}; 
\node at (-0.7,-1.7) {b)};
\draw[thick] (4.1,-5)  node[black,anchor=north]{$1$};
\draw[red,thick] (4.1,-5)--(4.1,-3.6)   ;
\draw[blue,thick,<-] (4.1,-3.6)--(4.1,-2.2) node[black,anchor=south]{$0$};
\draw[thick] (7.1,-5)  node[black,anchor=north]{$0$};
\draw[blue,dashed,thick] (7.1,-5)  -- (7.1,-3.6)  ;
\draw[red,dashed,thick,<-] (7.1,-3.6)  -- (7.1,-2.2) node[black,anchor=south]{$1$} ;
\end{tikzpicture}
}
\caption{Diagrammatic representation of cladogenetic state change 
in panel a) and anagenetic state change in panel b). The $0$ and $1$ 
labels on the figures represent trait marks. In a), cladogentic state 
change $0\to 0+0$ is given by two solid blue lines; change $1\to 1+1$ 
by two solid red lines; change $0\to 0+1$ by a solid blue and a solid red line; 
and change $1\to 0+1$ by a dashed blue and a dashed red line. In b), 
anagenetic state change $0\to 1$ is given by a solid blue-red line; and 
change $1\to 0$ by a dashed red-blue line.} 
\label{biss}
\end{figure}
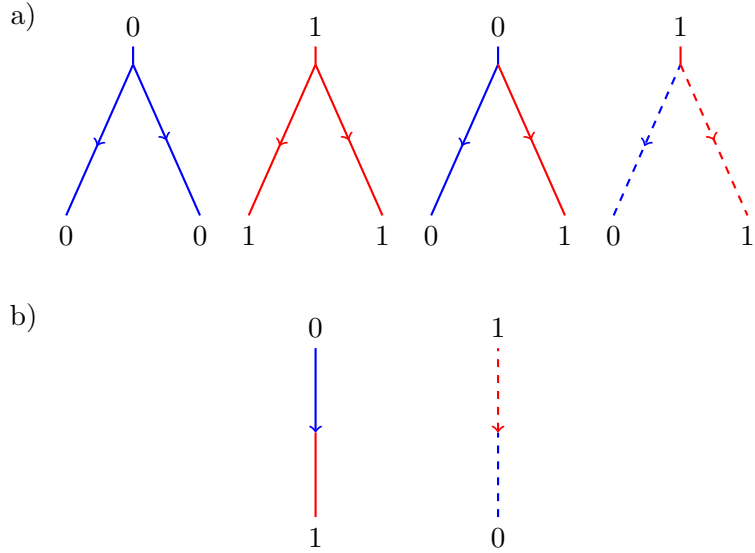

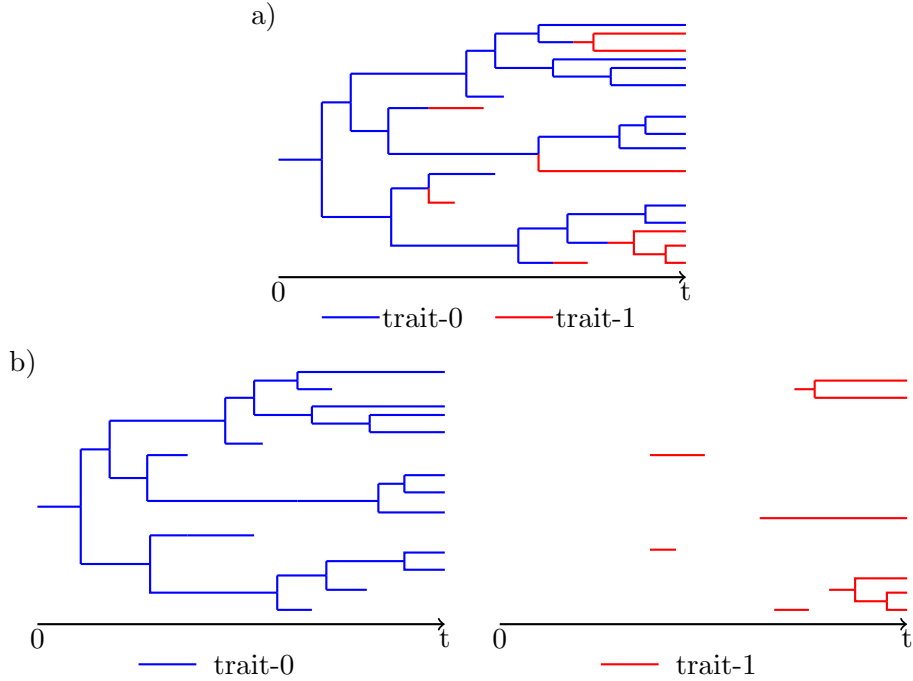
\begin{figure*}[!t]
\centerline{
\begin{minipage}[c]{0.42\linewidth}
\begin{tikzpicture}[scale=0.38]
\draw [->] [thick](-0.5,1.9) -- (13.6,1.9);
\node at (-1,11) {a)};
\node at (-.5,1.4) {0};
\node at (13.6,1.4) {t};
\draw [blue][thick] (1,.5) -- (3,.5);
\draw [red][thick] (7,.5) -- (9,.5);
\node at (4.5,.5) {trait-$0$};
\node at (10.5,.5) {trait-$1$};
\draw [blue][thick] (-0.5,6) -- (1,6);
\draw [blue] [thick] (1,6)--(1,8);
\draw [blue] [thick] (1,8)--(2,8);
\draw [blue] [thick] (2,8)--(2,9);
\draw [blue] [thick] (2,8)--(2,7);
\draw [blue] [thick] (2,9)--(6,9);
\draw [blue] [thick] (6,9)--(6,9.8);
\draw [blue] [thick] (2,7)--(3.3,7);
\draw [blue] [thick] (3.3,7.8)--(3.3,6.2);
\draw [blue] [thick] (3.3,6.2)--(8.5,6.2);
\draw [blue] [thick] (8.5,6.2)--(8.5,6.8);
\draw [blue] [thick](8.5,6.8)--(11.3,6.8);
\draw [blue] [thick] (11.3,6.8)--(11.3,7.2);
\draw [blue] [thick] (11.3,7.2)--(12.2,7.2);
\draw [blue] [thick] (12.2,7.2)--(12.2,7.5);
\draw [blue] [thick] (12.2,7.5)--(13.6,7.5);
\draw [red] [thick] (8.5,6.2)--(8.5,5.6);
\draw [red] [thick] (8.5,5.6)--(13.6,5.6);
\draw [blue] [thick] (12.2,7.2)--(12.2,6.9);
\draw [blue] [thick] (12.2,6.9)--(13.6,6.9);
\draw [blue] [thick] (11.3,6.8)--(11.3,6.4);
\draw [blue] [thick] (11.3,6.4)--(13.6,6.4);
\draw [blue] [thick] (3.3,7.8)--(4.7,7.8);
\draw [red] [thick] (4.7,7.8)--(6.6,7.8);
\draw [blue] [thick] (6,9)--(6,8.2);
\draw [blue] [thick] (6,9.8)--(7,9.8);
\draw [blue] [thick] (6,8.2)--(7.3,8.2);
\draw [blue] [thick] (7,9.8)--(7,10.4);
\draw [blue] [thick] (7,9.8)--(7,9.2);
\draw [blue] [thick] (7,10.4)--(8.5,10.4);
\draw [blue] [thick] (8.5,10.4)--(8.5,10.7);
\draw [blue] [thick] (8.5,10.4)--(8.5,10.1);
\draw [blue] [thick] (8.5,10.7)--(13.6,10.7);
\draw [blue] [thick] (8.5,10.1)--(9.7,10.1);
\draw [red] [thick] (9.7,10.1)--(10.4,10.1);
\draw [red] [thick] (10.4,10.1)--(10.4,10.4);
\draw [red] [thick] (10.4,10.4)--(13.6,10.4);
\draw [red] [thick] (10.4,10.1)--(10.4,10.1-.3);
\draw [red] [thick] (10.4,10.1-.3)--(13.6,10.1-.3);
\draw [blue] [thick] (7,9.2)--(9,9.2);
\draw [blue] [thick] (9,9.2)--(9,9.5);
\draw [blue] [thick] (9,9.5)--(13.6,9.5);
\draw [blue] [thick] (9,9.2)--(9,8.9);
\draw [blue] [thick] (9,8.9)--(11,8.9);
\draw [blue] [thick] (11,8.9)--(11,8.6);
\draw [blue] [thick] (11,8.6)--(13.6,8.6);
\draw [blue] [thick] (11,8.9)--(11,9.2);
\draw [blue] [thick] (11,9.2)--(13.6,9.2);
\draw [blue] [thick] (1,6)--(1,4);
\draw [blue] [thick] (1,4)--(3.4,4);
\draw [blue] [thick] (3.4,4)--(3.4,5);
\draw [blue] [thick] (3.4,5)--(4.7,5);
\draw [blue] [thick] (4.7,5)--(4.7,5.5);
\draw [blue] [thick] (4.7,5.5)--(7,5.5);
\draw [red] [thick] (4.7,5)--(4.7,4.5)--(5.6,4.5);
\draw [blue] [thick] (3.4,4)--(3.4,3)--(7.8,3);
\draw [blue] [thick] (7.8,3)--(7.8,3.6);
\draw [blue] [thick] (7.8,3)--(7.8,2.4);
\draw [blue] [thick] (7.8,2.4)--(9,2.4);
\draw [red] [thick] (9,2.4)--(10.2,2.4);
\draw [blue] [thick] (7.8,3.6)--(9.5,3.6);
\draw [blue] [thick] (9.5,3.6)--(9.5,4.1);
\draw [blue] [thick] (9.5,3.6)--(9.5,3.1);
\draw [blue] [thick] (9.5,4.1)--(12.2,4.1);
\draw [blue] [thick] (12.2,4.1)--(12.2,4.4)--(13.6,4.4);
\draw [blue] [thick] (12.2,4.1)--(12.2,3.8)--(13.6,3.8);
\draw [blue] [thick] (9.5,3.1)--(10.9,3.1);
\draw [red] [thick] (10.9,3.1)--(11.8,3.1);
\draw [red] [thick] (11.8,3.1)--(11.8,3.5)--(13.6,3.5);
\draw [red] [thick] (11.8,3.1)--(11.8,2.7)--(12.9,2.7);
\draw [red] [thick] (12.9,2.7)--(12.9,2.4)--(13.6,2.4);
\draw [red] [thick] (12.9,2.7)--(12.9,3)--(13.6,3);
\end{tikzpicture}
\end{minipage}}

\centerline{
\begin{minipage}[c]{0.87\linewidth}
\begin{tikzpicture}[scale=.38]
\draw [->] [thick](-0.5,1.9) -- (13.6,1.9);
\draw [->] [thick](-0.5+16,1.9) -- (13.6+16,1.9);
\node at (-1,11) {b)};
\node at (-.5,1.4) {0};
\node at (-.5+16,1.4) {0};
\node at (13.6,1.4) {t};
\node at (13.6+16,1.4) {t};
\draw [blue][thick] (3,.5) -- (5,.5);
\draw [red][thick] (8+11,.5) -- (10+11,.5);
\node at (7,.5) {trait-$0$};
\node at (12+11,.5) {trait-$1$};
\draw [blue][thick] (-0.5,6) -- (1,6);
\draw [blue] [thick] (1,6)--(1,8);
\draw [blue] [thick] (1,8)--(2,8);
\draw [blue] [thick] (2,8)--(2,9);
\draw [blue] [thick] (2,8)--(2,7);
\draw [blue] [thick] (2,9)--(6,9);
\draw [blue] [thick] (6,9)--(6,9.8);
\draw [blue] [thick] (2,7)--(3.3,7);
\draw [blue] [thick] (3.3,7.8)--(3.3,6.2);
\draw [blue] [thick] (3.3,6.2)--(8.5,6.2);
\draw [blue] [thick](8.5,6.2)--(11.3,6.2);
\draw [blue] [thick] (11.3,6.2)--(11.3,6.8);
\draw [blue] [thick] (11.3,6.8)--(12.2,6.8);
\draw [blue] [thick] (12.2,6.8)--(12.2,7.1);
\draw [blue] [thick] (12.2,7.1)--(13.6,7.1);
\draw [blue] [thick] (12.2,6.8)--(12.2,6.5);
\draw [blue] [thick] (12.2,6.5)--(13.6,6.5);
\draw [blue] [thick] (11.3,6.2)--(11.3,5.8);
\draw [blue] [thick] (11.3,5.8)--(13.6,5.8);
\draw [blue] [thick] (3.3,7.8)--(4.7,7.8);
\draw [blue] [thick] (6,9)--(6,8.2);
\draw [blue] [thick] (6,9.8)--(7,9.8);
\draw [blue] [thick] (6,8.2)--(7.3,8.2);
\draw [blue] [thick] (7,9.8)--(7,10.4);
\draw [blue] [thick] (7,9.8)--(7,9.2);
\draw [blue] [thick] (7,10.4)--(8.5,10.4);
\draw [blue] [thick] (8.5,10.4)--(8.5,10.7);
\draw [blue] [thick] (8.5,10.4)--(8.5,10.1);
\draw [blue] [thick] (8.5,10.7)--(13.6,10.7);
\draw [blue] [thick] (8.5,10.1)--(9.7,10.1);
\draw [blue] [thick] (7,9.2)--(9,9.2);
\draw [blue] [thick] (9,9.2)--(9,9.5);
\draw [blue] [thick] (9,9.5)--(13.6,9.5);
\draw [blue] [thick] (9,9.2)--(9,8.9);
\draw [blue] [thick] (9,8.9)--(11,8.9);
\draw [blue] [thick] (11,8.9)--(11,8.6);
\draw [blue] [thick] (11,8.6)--(13.6,8.6);
\draw [blue] [thick] (11,8.9)--(11,9.2);
\draw [blue] [thick] (11,9.2)--(13.6,9.2);
\draw [blue] [thick] (1,6)--(1,4);
\draw [blue] [thick] (1,4)--(3.4,4);
\draw [blue] [thick] (3.4,4)--(3.4,5);
\draw [blue] [thick] (3.4,5)--(4.7,5);
\draw [blue] [thick] (4.7,5)--(7,5);
\draw [blue] [thick] (3.4,4)--(3.4,3)--(7.8,3);
\draw [blue] [thick] (7.8,3)--(7.8,3.6);
\draw [blue] [thick] (7.8,3)--(7.8,2.4);
\draw [blue] [thick] (7.8,2.4)--(9,2.4);
\draw [blue] [thick] (7.8,3.6)--(9.5,3.6);
\draw [blue] [thick] (9.5,3.6)--(9.5,4.1);
\draw [blue] [thick] (9.5,3.6)--(9.5,3.1);
\draw [blue] [thick] (9.5,4.1)--(12.2,4.1);
\draw [blue] [thick] (12.2,4.1)--(12.2,4.4)--(13.6,4.4);
\draw [blue] [thick] (12.2,4.1)--(12.2,3.8)--(13.6,3.8);
\draw [blue] [thick] (9.5,3.1)--(10.9,3.1);
\draw [red] [thick] (8.5+16,5.6)--(13.6+16,5.6);
\draw [red] [thick] (4.7+16,7.8)--(6.6+16,7.8);
\draw [red] [thick] (9.7+16,10.1)--(10.4+16,10.1);
\draw [red] [thick] (10.4+16,10.1)--(10.4+16,10.4);
\draw [red] [thick] (10.4+16,10.4)--(13.6+16,10.4);
\draw [red] [thick] (10.4+16,10.1)--(10.4+16,10.1-.3);
\draw [red] [thick] (10.4+16,10.1-.3)--(13.6+16,10.1-.3);
\draw [red] [thick] (4.7+16,4.5)--(5.6+16,4.5);
\draw [red] [thick] (9+16,2.4)--(10.2+16,2.4);
\draw [red] [thick] (10.9+16,3.1)--(11.8+16,3.1);
\draw [red] [thick] (11.8+16,3.1)--(11.8+16,3.5)--(13.6+16,3.5);
\draw [red] [thick] (11.8+16,3.1)--(11.8+16,2.7)--(12.9+16,2.7);
\draw [red] [thick] (12.9+16,2.7)--(12.9+16,2.4)--(13.6+16,2.4);
\draw [red] [thick] (12.9+16,2.7)--(12.9+16,3)--(13.6+16,3);
\end{tikzpicture}
\end{minipage}}
\caption{Diagram showing in panel a): the rooted tree $\mathcal T=\mathcal T^0\cup \mathcal T^1$, with trait-$0$
  species colored blue, and trait-$1$ species colored red; and in panel b): the rooted, type-$0$ tree, $\mathcal T^0$, and
  edges of type-$1$ trees, $\mathcal T^1$, plotted separately, in blue and red, respectively.}
\label{fig:treesplit}
\end{figure*}

We consider a binary trait on a family tree of species, starting with
a single species of known trait at the root of the tree.  Each species
in the tree, throughout its lifetime, carries trait value $0$ or $1$.
The species family evolves as a branching tree with births of new
species and extinction of existing species.  The tree runs in
continuous time over an interval $[0,t]$, where $t$ is the time span
from the arrival of the first species at $t=0$, up to the time of
observation -- today.  The ancestral species carries trait $0$.  The
speciation process in the model is the simplest possible where new
species arise instantly adding a new node to the tree, either as a
branch point or a transition from an earlier species.  Each birth
event either replaces one species with two new ones and a combination
of traits representing cladogenetic state change (solid lines in
Fig. \ref{biss}a), or replaces one species with one new, representing
anagenetic state change (solid line in Fig. \ref{biss}b).  The number
of species and their traits change according to an integer-valued,
two-type, continuous time Markov branching process, where we use type
and trait interchangebly (\cite{athreya_ney};
\cite{taylor_karlin}).  The Markov intensities for birth events are
$\lambda_0$ for a change $0\to 0+0$, $\lambda_1$ for $1\to 1+1$,
$p\delta$ for $0\to 0+1$ and $(1-p)\delta$ for $0\to 1$. Here,
$\lambda_0$, $\lambda_1$, $\delta$ are non-negative jump rates and $p$, $0\le p\le 1$,
is the probability of
cladogenetic change of states from type-$0$ to type-$1$ species.
 The case $p=0$ is entirely anagenetic
change and the case $p=1$ entirely cladogenetic.  Each species is
exposed to exctinction with intensities $\mu_0$ for type-$0$ and
$\mu_1$ for type-$1$.  The resulting species family is described by a
rooted tree $\mathcal T$ with vertices given by the birth or
extinction of species. The edges are marked by $0$ or $1$ recording
the trait of a species and the edge lengths represent the species
lifetime.  Based on the edge marks, the full tree is composed of
disjoint parts
\[
\mathcal T=\mathcal T^0\cup \mathcal T^1,
\]
where $\mathcal T^\mathrm{0}$ is connected with root at $t=0$ and
$\mathcal T^\mathrm{1}$ consists of all edges of type-$1$.  
Figure \ref{fig:treesplit}a shows a tree, $\mathcal T$ restricted to the
interval $[0,t]$ with species of trait-$0$ plotted in blue and species
of trait-$1$ plotted in red. Figure \ref{fig:treesplit}b
shows the graphs $\mathcal T^0$ and $\mathcal T^1$ separately, again cut off at time
$t$. 

\begin{center}
\begin{table}
\caption{Description of important parameters that are used in this paper 
\label{t1}
}}
{\tabcolsep=4pt
\begin{tabular}{@{\extracolsep{\fill}}ll@{}}
\hline
Parameter & Description \\\hline\\
$K_t$ & number of type-$0$ species\\
$L_t$ &  number of type-$1$ species\\
$\lambda_0=\lambda$ &  rate of speciation of  type-$0$ species\\ 
$\lambda_1$ &  rate of speciation of  type-$1$ species \\ 
$\mu_0$ & rate of extinction of type-$0$ species \\
$\mu_1$ & rate of extinction of type-$1$ species\\
$r_0$ & rate of diversification of type-$0$ species ($\lambda_0-\mu_0$)\\
$r_1= \gamma_1$ & rate of diversification of type-$1$ species  ($\lambda_1-\mu_1$)\\
$p\delta$ & rate of cladogenetic change from type-$0$ to type-$1$\\
$(1-p)\delta$ & rate of anagenetic change from type-$0$ to type-$1$\\
$\delta$ & total rate of state change from type-$0$ to type-$1$\\
$\mu$ & rate of removal of type-$0$ species ($\mu_0+(1-p)\delta$)\\
$\gamma_0$ & $\lambda-\mu=r_0-(1-p)\delta$\\
$\mathcal T$ & complete species tree\\
$\mathcal T^0$ &  trait-$0$ component in $\mathcal T$\\ 
$\mathcal T^1$ & trait-$1$ component in $\mathcal T$\\ 
$\mathcal T_t$ & complete, reduced species tree at time $t$\\ 
$(\mathcal T_t)^0$ & trait-$0$ component in $\mathcal T_t$\\ 
$(\mathcal T_t)^1$ & trait-$1$ component in $\mathcal T_t$\\ 
$(\mathcal T^0)_t$ & trait-$0$ species tree, reduced at time $t$\\ 
$(\mathcal T^1)_t$ & trait-$1$ species tree, reduced at time $t$\\ 
$A_t$ & total branch length of $(\mathcal T^0)_t$\\ 
$A'_t$ & total branch length of $(\mathcal T_t)^0$\\
$B_t$ & total branch length of $(\mathcal T^1)_t=(\mathcal T_t)^1$\\
$\widetilde K_{s,t}$ & number of trait-$0$ species  in $(\mathcal T^0)_t$ at time $s$\\
$\widetilde K'_{s,t}$ & number of trait-$0$ species in $(\mathcal T_t)^0$ at time $s$\\
$\widetilde L_{s,t}$ & number of trait-$1$ species in $(\mathcal T_t)^1$ at time $s$\\
$C_t$ & number of clusters of trait-$1$ species at $t$\\
$L/C$ & ratio of $E(L_t|K_t>0)$ to $E(C_t|K_t>0)$\\
$r_0(s)$ & expected time spent as trait-$0$ on a\\
&   trait-$0$/trait-$1$ branch of length $s$\\
$dN/dS$ & normalized ratio of nonsynonymous to\\
&  synonymous substitutions\\
$\omega_0$ & $dN/dS$ in type-$0$ species\\
$\omega_1$ & $dN/dS$ in type-$1$ species\\
$z$ &  $\frac{\mu_1}{\mu_0}=\frac{\omega_1}{\omega_0}$\\
$T_\mathrm{tot}$ & branch length of the outcrosser-selfer species tree\\ 
$T^{(0)}_\mathrm{tot}$ & branch length of the outcrosser species tree\\
$T^{(1)}_\mathrm{tot}$ & branch length of the selfer species tree\\
  \hline
\end{tabular}}
{
\end{table}
\end{center}

Putting $K_t=$ the number of type-$0$ species and $L_t=$ the 
number of type-1 species, $K_t+L_t$ is the total
number of species at time $t$, and
\begin{equation}\label{branchingprocess}
X_t=(K_t,L_t),\quad t\ge 0,
\end{equation}
is a two-type continuous time Markov branching process with branching 
rates
\begin{equation}\label{branchingrates}
(k,\ell)\mapsto \left\{
\begin{array}{cc}
(k+1,\ell) & \lambda_0 k\\
(k-1,\ell+1) & (1-p)\delta k\\
(k-1,\ell) & \mu_0 k\\
(k,\ell+1) & p\delta k+\lambda_1 \ell\\
(k,\ell-1) & \mu_1 \ell.
\end{array}
\right.
\end{equation}
We have $X_0=(1,0)$, that is $K_0=1$, $L_0=0$.  Of
course, $K_t$, the number of species in the sub-tree $\mathcal T^0$, is an
ordinary one-type continuous time branching process with parameters
$(\lambda,\mu)$, where $\lambda=\lambda_0$ is the rate of binary
splitting of the $0$-trait and $\mu=\mu_0+(1-p)\delta$ is the rate of
removal of trait-$0$ species. The exact distribution of $X_t$ is known
for the anagenetic case $p=0$ in terms of generating functions 
\cite{antal_krapivsky}. The net diversification rates in the model, $r_0$ and
$r_1$ for the two traits, are
\begin{equation}\label{}
r_0=\lambda_0-\mu_0,\quad r_1=\lambda_1-\mu_1
\end{equation}
and the eigenvalues of the mean offspring matrix are given by 
\begin{equation}\label{netrepro}
\gamma_0=\lambda-\mu=r_0-(1-p)\delta,\quad \gamma_1=r_1.
\end{equation}
A list of all important parameters, used in this paper, is given in
Table \ref{t1}.  Appendix 1 summarizes the mathematical properties of
the two-type branching process $X_t$.  The more general model of
two-sided transitions, where type-$1$ species may generate species of
type-$0$ at birth (dashed lines in Fig. \ref{biss}a and
Fig. \ref{biss}b) is not discussed further in this work.  Our approach
does not seem to adapt easily to this case, where $X_t$ is a more
general two-type branching process.

\subsection*{The Mutation Process}

Mutation events occur randomly, according to a fixed Poisson molecular clock
of evolution, and a resulting Poisson intensity $\theta>0$ of mutations
per time unit and ``gene'', the same for all species.  The actual
marks of mutation, such as nucleotide substitutions, codon
substitutions, e.t.c., will be called `fixed mutations' for short. The rate
of fixed mutations depend on the trait. Indeed, letting 
$\omega_0$ and $\omega_1$ be the trait-dependent scaled fixation rates,
the fixed mutations accumulate as a Poisson process $\mathcal{N}$ running along
all branches of the species tree, with intensity $\theta\omega_0$ on
$\mathcal T^0$ and $\theta\omega_1$ on $\mathcal T^1$. The ordering
\begin{equation}\nonumber
 \omega_0<\omega_1<1,
\end{equation}
indicates that both traits are under negative selection, with the
efficacy of deleterious selection higher in trait-$0$.  Restricting to
those fixed mutations that are visible at $t$, leads to investigating
the so called `reduced branching tree', consisting of the sub-tree of
species having at least one descendant at $t$.  Let $\mathcal{N}^{(i)}_t$ be the
number of fixed mutations of type-$i$, $i=0,1$, observable at $t$. With 

\begin{align*}
A_t&=\mbox{total life time of $0$-species with descendants at $t$},\\ 
B_t&=\mbox{total life time of $1$-species with descendants at $t$},
\end{align*}

\noindent it follows that $\mathcal{N}^{(0)}_t$ and $\mathcal{N}^{(1)}_t$ are Poisson random
variables modulated by the random intensitities $\theta \omega_0 A_t$
and $\theta \omega_1 B_t$, respectively.  In particular, 
\[
E[\mathcal{N}^{(0)}_t]=\theta \omega_0 E[A_t],\quad
E[\mathcal{N}^{(1)}_t]=\theta \omega_1 E[B_t].
\]
The purpose of the next section is to find the expected values of
$A_t$ and $B_t$, and to relate these quantities to model parameters.
Under the dead-end hypothesis, accumulation of deleterious mutations
directly affects the long-term survival of species or is, at least, a
signature of reduced selection efficacy that can drive species towards
extinction. Hence, we propose to use a control parameter $c>0$, 
and the additional modeling assumption
\begin{equation}\label{c}
\mu_0=c\omega_0,\quad \mu_1=c\omega_1,
\end{equation}
as a link between the mutation processes and the species tree
development. More realistic parameterizations could be considered, but
would complicate the treatment without changing the rationale and
bringing new insight.

\section{ Analyzing the Reduced Tree}

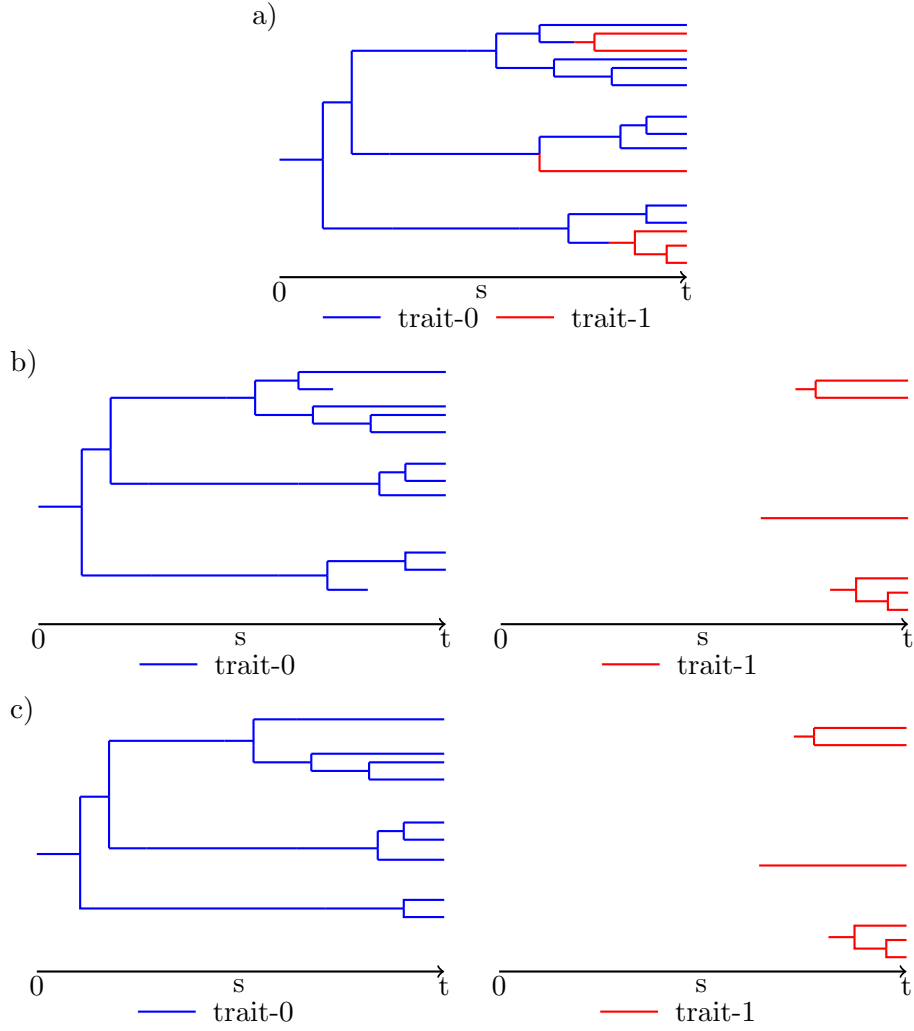
\begin{figure*}[!t]
\centerline{
\begin{minipage}[c]{0.42\linewidth}
\begin{tikzpicture}[scale=0.38]
\draw [->] [thick](-0.5,1.9) -- (13.6,1.9);
\node at (-1,11) {a)};
\node at (-.5,1.4) {0};
\node at (6.5,1.4) {s};
\node at (13.6,1.4) {t};
\draw [blue][thick] (1,.5) -- (3,.5);
\draw [red][thick] (7,.5) -- (9,.5);
\node at (5,.5) {trait-$0$};
\node at (11,.5) {trait-$1$};
\draw [blue][thick] (-0.5,6) -- (1,6);
\draw [blue] [thick] (1,6)--(1,8);
\draw [blue] [thick] (1,8)--(2,8);
\draw [blue] [thick] (2,8)--(2,9.8);
\draw [blue] [thick] (2,8)--(2,6.2);
\draw [blue] [thick] (2,9.8)--(6,9.8);
\draw [blue] [thick] (2,6.2)--(3.3,6.2);
\draw [blue] [thick] (3.3,6.2)--(8.5,6.2);
\draw [blue] [thick] (8.5,6.2)--(8.5,6.8);
\draw [blue] [thick](8.5,6.8)--(11.3,6.8);
\draw [blue] [thick] (11.3,6.8)--(11.3,7.2);
\draw [blue] [thick] (11.3,7.2)--(12.2,7.2);
\draw [blue] [thick] (12.2,7.2)--(12.2,7.5);
\draw [blue] [thick] (12.2,7.5)--(13.6,7.5);
\draw [red] [thick] (8.5,6.2)--(8.5,5.6);
\draw [red] [thick] (8.5,5.6)--(13.6,5.6);
\draw [blue] [thick] (12.2,7.2)--(12.2,6.9);
\draw [blue] [thick] (12.2,6.9)--(13.6,6.9);
\draw [blue] [thick] (11.3,6.8)--(11.3,6.4);
\draw [blue] [thick] (11.3,6.4)--(13.6,6.4);
\draw [blue] [thick] (6,9.8)--(7,9.8);
\draw [blue] [thick] (7,9.8)--(7,10.4);
\draw [blue] [thick] (7,9.8)--(7,9.2);
\draw [blue] [thick] (7,10.4)--(8.5,10.4);
\draw [blue] [thick] (8.5,10.4)--(8.5,10.7);
\draw [blue] [thick] (8.5,10.4)--(8.5,10.1);
\draw [blue] [thick] (8.5,10.7)--(13.6,10.7);
\draw [blue] [thick] (8.5,10.1)--(9.7,10.1);
\draw [red] [thick] (9.7,10.1)--(10.4,10.1);
\draw [red] [thick] (10.4,10.1)--(10.4,10.4);
\draw [red] [thick] (10.4,10.4)--(13.6,10.4);
\draw [red] [thick] (10.4,10.1)--(10.4,10.1-.3);
\draw [red] [thick] (10.4,10.1-.3)--(13.6,10.1-.3);
\draw [blue] [thick] (7,9.2)--(9,9.2);
\draw [blue] [thick] (9,9.2)--(9,9.5);
\draw [blue] [thick] (9,9.5)--(13.6,9.5);
\draw [blue] [thick] (9,9.2)--(9,8.9);
\draw [blue] [thick] (9,8.9)--(11,8.9);
\draw [blue] [thick] (11,8.9)--(11,8.6);
\draw [blue] [thick] (11,8.6)--(13.6,8.6);
\draw [blue] [thick] (11,8.9)--(11,9.2);
\draw [blue] [thick] (11,9.2)--(13.6,9.2);
\draw [blue] [thick] (1,6)--(1,4);
\draw [blue] [thick] (1,3.6)--(3.4,3.6);
\draw [blue] [thick] (1,3.6)--(1,4);
\draw [blue] [thick]  (3.4,3.6)--(7.8,3.6);
\draw [blue] [thick] (7.8,3.6)--(9.5,3.6);
\draw [blue] [thick] (9.5,3.6)--(9.5,4.1);
\draw [blue] [thick] (9.5,3.6)--(9.5,3.1);
\draw [blue] [thick] (9.5,4.1)--(12.2,4.1);
\draw [blue] [thick] (12.2,4.1)--(12.2,4.4)--(13.6,4.4);
\draw [blue] [thick] (12.2,4.1)--(12.2,3.8)--(13.6,3.8);
\draw [blue] [thick] (9.5,3.1)--(10.9,3.1);
\draw [red] [thick] (10.9,3.1)--(11.8,3.1);
\draw [red] [thick] (11.8,3.1)--(11.8,3.5)--(13.6,3.5);
\draw [red] [thick] (11.8,3.1)--(11.8,2.7)--(12.9,2.7);
\draw [red] [thick] (12.9,2.7)--(12.9,2.4)--(13.6,2.4);
\draw [red] [thick] (12.9,2.7)--(12.9,3)--(13.6,3);
\end{tikzpicture}
\end{minipage}}

\centerline{
\begin{minipage}[c]{0.87\linewidth}
\begin{tikzpicture}[scale=.38]
\draw [->] [thick](-0.5,1.9) -- (13.6,1.9);
\draw [->] [thick](-0.5+16,1.9) -- (13.6+16,1.9);
\node at (-1,11) {b)};
\node at (-.5,1.4) {0};
\node at (-.5+16,1.4) {0};
\node at (6.5,1.4) {s};
\node at (6.5+16,1.4) {s};
\node at (13.6,1.4) {t};
\node at (13.6+16,1.4) {t};
\draw [blue][thick] (3,.5) -- (5,.5);
\draw [red][thick] (8+11,.5) -- (10+11,.5);
\node at (7,.5) {trait-$0$};
\node at (12+11,.5) {trait-$1$};
\draw [blue][thick] (-0.5,6) -- (1,6);
\draw [blue] [thick] (1,6)--(1,8);
\draw [blue] [thick] (1,8)--(2,8);
\draw [blue] [thick] (2,8)--(2,9.8);
\draw [blue] [thick] (2,8)--(2,6.8);
\draw [blue] [thick] (2,9.8)--(6,9.8);
\draw [blue] [thick] (2,6.8)--(3.3,6.8);
\draw [blue] [thick] (3.3,6.8)--(8.5,6.8);
\draw [blue] [thick](8.5,6.8)--(11.3,6.8);
\draw [blue] [thick] (11.3,6.8)--(11.3,7.2);
\draw [blue] [thick] (11.3,7.2)--(12.2,7.2);
\draw [blue] [thick] (12.2,7.2)--(12.2,7.5);
\draw [blue] [thick] (12.2,7.5)--(13.6,7.5);
\draw [blue] [thick] (12.2,7.2)--(12.2,6.9);
\draw [blue] [thick] (12.2,6.9)--(13.6,6.9);
\draw [blue] [thick] (11.3,6.8)--(11.3,6.4);
\draw [blue] [thick] (11.3,6.4)--(13.6,6.4);
\draw [blue] [thick] (6,9.8)--(7,9.8);
\draw [blue] [thick] (7,9.8)--(7,10.4);
\draw [blue] [thick] (7,9.8)--(7,9.2);
\draw [blue] [thick] (7,10.4)--(8.5,10.4);
\draw [blue] [thick] (8.5,10.4)--(8.5,10.7);
\draw [blue] [thick] (8.5,10.4)--(8.5,10.1);
\draw [blue] [thick] (8.5,10.7)--(13.6,10.7);
\draw [blue] [thick] (8.5,10.1)--(9.7,10.1);
\draw [blue] [thick] (7,9.2)--(9,9.2);
\draw [blue] [thick] (9,9.2)--(9,9.5);
\draw [blue] [thick] (9,9.5)--(13.6,9.5);
\draw [blue] [thick] (9,9.2)--(9,8.9);
\draw [blue] [thick] (9,8.9)--(11,8.9);
\draw [blue] [thick] (11,8.9)--(11,8.6);
\draw [blue] [thick] (11,8.6)--(13.6,8.6);
\draw [blue] [thick] (11,8.9)--(11,9.2);
\draw [blue] [thick] (11,9.2)--(13.6,9.2);
\draw [blue] [thick] (1,6)--(1,4);
\draw [blue] [thick] (1,3.6)--(3.4,3.6);
\draw [blue] [thick] (1,3.6)--(1,4);
\draw [blue] [thick]  (3.4,3.6)--(7.8,3.6);
\draw [blue] [thick] (7.8,3.6)--(9.5,3.6);
\draw [blue] [thick] (9.5,3.6)--(9.5,4.1);
\draw [blue] [thick] (9.5,3.6)--(9.5,3.1);
\draw [blue] [thick] (9.5,4.1)--(12.2,4.1);
\draw [blue] [thick] (12.2,4.1)--(12.2,4.4)--(13.6,4.4);
\draw [blue] [thick] (12.2,4.1)--(12.2,3.8)--(13.6,3.8);
\draw [blue] [thick] (9.5,3.1)--(10.9,3.1);
\draw [red] [thick] (8.5+16,5.6)--(13.6+16,5.6);
\draw [red] [thick] (9.7+16,10.1)--(10.4+16,10.1);
\draw [red] [thick] (10.4+16,10.1)--(10.4+16,10.4);
\draw [red] [thick] (10.4+16,10.4)--(13.6+16,10.4);
\draw [red] [thick] (10.4+16,10.1)--(10.4+16,10.1-.3);
\draw [red] [thick] (10.4+16,10.1-.3)--(13.6+16,10.1-.3);
\draw [red] [thick] (10.9+16,3.1)--(11.8+16,3.1);
\draw [red] [thick] (11.8+16,3.1)--(11.8+16,3.5)--(13.6+16,3.5);
\draw [red] [thick] (11.8+16,3.1)--(11.8+16,2.7)--(12.9+16,2.7);
\draw [red] [thick] (12.9+16,2.7)--(12.9+16,2.4)--(13.6+16,2.4);
\draw [red] [thick] (12.9+16,2.7)--(12.9+16,3)--(13.6+16,3);
\end{tikzpicture}
\end{minipage}}   

\centerline{
\begin{minipage}[c]{0.87\linewidth}
\begin{tikzpicture}[scale=.38]
\draw [->] [thick](-0.5,1.9) -- (13.6,1.9);
\draw [->] [thick](-0.5+16,1.9) -- (13.6+16,1.9);
\node at (-1,11) {c)};
\node at (-.5,1.4) {0};
\node at (-.5+16,1.4) {0};
\node at (6.5,1.4) {s};
\node at (6.5+16,1.4) {s};
\node at (13.6,1.4) {t};
\node at (13.6+16,1.4) {t};
\draw [blue][thick] (3,.5) -- (5,.5);
\draw [red][thick] (8+11,.5) -- (10+11,.5);
\node at (7,.5) {trait-$0$};
\node at (12+11,.5) {trait-$1$};
\draw [blue][thick] (-0.5,6) -- (1,6);
\draw [blue] [thick] (1,6)--(1,8);
\draw [blue] [thick] (1,8)--(2,8);
\draw [blue] [thick] (2,8)--(2,9.95);
\draw [blue] [thick] (2,8)--(2,6.2);
\draw [blue] [thick] (2,9.95)--(6,9.95);
\draw [blue] [thick] (2,6.2)--(3.3,6.2);
\draw [blue] [thick] (3.3,6.2)--(8.5,6.2);
\draw [blue] [thick](8.5,6.2)--(11.3,6.2);
\draw [blue] [thick] (11.3,6.2)--(11.3,6.8);
\draw [blue] [thick] (11.3,6.8)--(12.2,6.8);
\draw [blue] [thick] (12.2,6.8)--(12.2,7.1);
\draw [blue] [thick] (12.2,7.1)--(13.6,7.1);
\draw [blue] [thick] (12.2,6.8)--(12.2,6.5);
\draw [blue] [thick] (12.2,6.5)--(13.6,6.5);
\draw [blue] [thick] (11.3,6.2)--(11.3,5.8);
\draw [blue] [thick] (11.3,5.8)--(13.6,5.8);
\draw [blue] [thick] (6,9.95)--(7,9.95);
\draw [blue] [thick] (7,9.8)--(7,10.7);
\draw [blue] [thick] (7,9.8)--(7,9.2);
\draw [blue] [thick] (7,10.7)--(8.5,10.7);
\draw [blue] [thick] (8.5,10.7)--(13.6,10.7);
\draw [blue] [thick] (7,9.2)--(9,9.2);
\draw [blue] [thick] (9,9.2)--(9,9.5);
\draw [blue] [thick] (9,9.5)--(13.6,9.5);
\draw [blue] [thick] (9,9.2)--(9,8.9);
\draw [blue] [thick] (9,8.9)--(11,8.9);
\draw [blue] [thick] (11,8.9)--(11,8.6);
\draw [blue] [thick] (11,8.6)--(13.6,8.6);
\draw [blue] [thick] (11,8.9)--(11,9.2);
\draw [blue] [thick] (11,9.2)--(13.6,9.2);
\draw [blue] [thick] (1,6)--(1,4.1)--(9.5,4.1);
\draw [blue] [thick] (9.5,4.1)--(12.2,4.1);
\draw [blue] [thick] (12.2,4.1)--(12.2,4.4)--(13.6,4.4);
\draw [blue] [thick] (12.2,4.1)--(12.2,3.8)--(13.6,3.8);
\draw [red] [thick] (8.5+16,5.6)--(13.6+16,5.6);
\draw [red] [thick] (9.7+16,10.1)--(10.4+16,10.1);
\draw [red] [thick] (10.4+16,10.1)--(10.4+16,10.4);
\draw [red] [thick] (10.4+16,10.4)--(13.6+16,10.4);
\draw [red] [thick] (10.4+16,10.1)--(10.4+16,10.1-.3);
\draw [red] [thick] (10.4+16,10.1-.3)--(13.6+16,10.1-.3);
\draw [red] [thick] (10.9+16,3.1)--(11.8+16,3.1);
\draw [red] [thick] (11.8+16,3.1)--(11.8+16,3.5)--(13.6+16,3.5);
\draw [red] [thick] (11.8+16,3.1)--(11.8+16,2.7)--(12.9+16,2.7);
\draw [red] [thick] (12.9+16,2.7)--(12.9+16,2.4)--(13.6+16,2.4);
\draw [red] [thick] (12.9+16,2.7)--(12.9+16,3)--(13.6+16,3);
\end{tikzpicture}
\end{minipage}}
\caption{Diagrammatic representation of reduced species trees. 
Panel a) shows the reduced tree $\mathcal T_t$, with trait-$0$
  species in blue and trait-$1$ species in red.
The trees in panel b) are obtained by splitting $\mathcal T_t$ 
  into disjoint parts, that is, $(\mathcal T_t)^0$ and 
  $(\mathcal T_t)^1$, which are plotted separately, in blue and red, respectively. 
  The figures in panel c) show the rooted trait-$0$ tree, $(\mathcal T^0)_t$, in blue
  and the trait-$1$ edges, $(\mathcal T^1)_t$, in red, which are are obtained if the full tree $\mathcal T$ 
  (from Fig. \ref{fig:treesplit}a) is first split into type-$0$ and type-$1$ disjoint parts and then reduced 
  at $t$. In all the three panels, $t\ge 0$ and $0\le s\le t$.  
  }   
\label{fig:reducesplit}
\end{figure*}

The reduced species tree is what remains if we fix a time point
and remove all branches of the full species tree, $\mathcal T$, that are extinct at
that time.  More formally, we denote by $\mathcal T_t$, $t\ge 0$, the sequence
of the reduced species trees, defined for each fixed $t$. 
$\mathcal T_t$ is pruned of any extinct species and hence composed of only those
branches which exist at time $t$ (Fig. \ref{fig:reducesplit}a). In
case, all species are extinct at $t$, the reduced tree is empty. The
species traits in the original tree provide a record of types for the reduced
tree, and hence $\mathcal T_t$ splits up into disjoint 
trees
\[
\mathcal T_t=(\mathcal T_t)^0\cup (\mathcal T_t)^1,
\]
where $(\mathcal T_t)^0$ is a single connected tree and
$(\mathcal T_t)^1$ may consist of several components . We observe,
however, that 
\[
(\mathcal T_t)^0\supseteq (\mathcal T^0)_t \quad \mbox{and} \quad
(\mathcal T_t)^1= (\mathcal T^1)_t,
\]
as shown in Figure \ref{fig:reducesplit}b and Figure \ref{fig:reducesplit}c. 
Indeed, for $0\le s\le t$, letting $\widetilde K_{s,t}$ be the size at
time $s$ of the tree $(\mathcal T^\mathrm{0})_t$ obtained by first splitting
and then reducing the full tree, and $\widetilde K'_{s,t}$ be the size
at $s$ of the corresponding tree $(\mathcal T_t)^\mathrm{0}$ obtained by
first reducing and then splitting, one has
\begin{align*}
\widetilde K_{s,t} &=\mbox{number of $0$-species at $s$
with at least one descendant $0$-species at $t$},\\
\widetilde K'_{s,t}&=\mbox{number of $0$-species at $s$
with at least one descendant $0$- or $1$-species at $t$},
\end{align*}
so that $\widetilde K_{s,t}\le \widetilde K'_{s,t}$.  For trait-$1$
species, the order of splitting and reducing the tree has no effect and
we write $\widetilde L_{s,t}$ for the total number of $1$-species at
time $s$ in the reduced tree $(\mathcal T_t)^1=(\mathcal T^1)_t$, consisting of the
collection of reduced trees generated by any existing $1$-species at
$t$. This collection of branches could be empty, consist of a single
reduced tree of siblings, or be composed of a cluster of disconnected
trees with at least one descendant, each at time $t$.  Clearly,
$\widetilde K_{t,t}=K_t$ and $\widetilde L_{t,t}=L_t$. When
$|X_t|=K_t+L_t=0$, the reduced tree is void and $\widetilde
K_{s,t}=\widetilde L_{s,t}=0$, where $0\le s\le t$.

The total branch length of $(\mathcal T^0)_t$ is given by
\begin{equation}\label{totalbranchout}
A_t=\int_0^t \widetilde K_{s,t}\,ds.
\end{equation} 
Let  
\begin{equation}\label{totalbranchoutplus}
A_t'=\int_0^t \widetilde K'_{s,t}\,ds
\end{equation} 
be the total branch length of $(\mathcal T_t)^0$, so that $A_t'\ge A_t$ almost
surely. The number of fixations
present at $t$, which originate from a mutation in a species of type-$0$,
 during the time interval $(0,t)$, is a Poisson variable with
stochastic intensity $\theta\omega_0\, A'_t$.  

\noindent Similarly, 
the total branch length of trait-$1$ species is given by
\begin{equation}\label{totalbranchself}
B_t=\int_0^t \widetilde L_{s,t}\,ds.
\end{equation}

Next, two existing approaches towards analyzing the reduced tree are
reviewed, unified, and extended.  One approach starts from the full
species tree conditioned to be non-empty at $t$, and extracts the size
of the reduced tree by probabilistic thinning (\cite{kendall}; \cite{nee_etal}). 
The other approach begins with a given size of the tree at $t$
and traces backwards in time to find the relevant bifurcation times
(\cite{thompson}; \cite{gernhard}).  The terminology of reduced trees used here,
is well established in the theory of branching processes.  Other
options are reconstructed tree or reconstructed evolutionary process,
being aware that statistically oriented phylogeneticists might use other
kinds of reconstructed trees.

\subsection*{The Trait-0 Tree Conditioned on Non-Extinction}

To determine the expected total branch length in the reduced tree
$(\mathcal T_t)^\mathrm{0}$, we recall the conditional expectations
of the branching process $K_t$, and the reduced branching process
$\widetilde K_{s,t}$ (see e.g., \cite{kendall} and \cite{nee_etal}).  With
$\lambda$, $\mu$, and $\gamma_0$ as in (\ref{netrepro}), and
restricting to $\gamma_0\not=0$, put
\[
p_0(t)=\frac{\mu(1-e^{-\gamma_0 t})}{\lambda-\mu e^{-\gamma_0t}}, \;\;\;\;
u_t=\frac{\lambda p_0(t)}{\mu}=\frac{\lambda(1-e^{-\gamma_0
    t})}{\lambda-\mu e^{-\gamma_0 t}},
\]
and
\[
v_{s,t}=\frac{\lambda(1-e^{-\gamma_0 s})}{\lambda-\mu e^{-\gamma_0 t}}.
\]
First of all,
\[
E(K_t|K_t>0)=\frac{1}{1-u_t} \quad \mbox{and} \quad E(K_t)=e^{\gamma_0 t}.
\]
More generally, for $s\le t$,
\begin{equation}\label{expectedK_s}
E(K_s|K_t>0)=\frac{1}{1-u_s}+\frac{u_sp_0(t-s)}{1-u_sp_0(t-s)},
\end{equation}
and
\begin{equation}\label{condmeanreduced}
E(\widetilde K_{s,t}|K_t>0)=\frac{1}{1-v_{s,t}}, \quad
E(\widetilde K_{s,t})=\frac{1-p_0(t)}{1-v_{s,t}}.
\end{equation}
Thus, 
\begin{align}
\label{dndsweightout}
E(A_t|K_t>0)=\int_0^t E(\widetilde K_{s,t}|K_t>0)\,ds
=\int_0^t \frac{\lambda e^{\gamma_0 t}-\mu}{\lambda e^{\gamma_0(t-s)}-\mu}\,ds.  
\end{align}
Evaluating the above integral, we obtain
\begin{equation}\nonumber
E(A_t|K_t>0)=\frac{\mu-\lambda e^{\gamma_0 t}}{\mu \gamma_0}
\log\Big(\frac{\lambda-\mu}{\lambda-\mu e^{-\gamma_0 t}}\Big).
\end{equation}
These relations simplify for the critical case
$\gamma_0=\lambda-\mu=0$, for example
 \[
E(A_t|K_t>0)
=\int_0^t \frac{1+\lambda t}{1+\lambda(t-s)}\,ds=\frac{1+\lambda
t}{\lambda}\log(1+\lambda t).
\]

\subsection*{The Reduced Trait-1 Species Tree}

The trait-$1$ species tree is composed of a collection of branches,
injected at random times and locations on top of the initially
existing trait-$0$ tree. The total intensity, at which species of trait-$1$ 
enter the tree at any time $s\ge 0$, is $\delta K_s$, hence
proportional to the current number of $0$-traits, $K_s$, in the system.
A new trait-$1$ species is the result of cladogenetic
splitting with probability $p$ and of anagenetic transition with
probability $1-p$. Each new $1$-species potentially initiates a sub-tree,
which preserves its trait during the subsequent path to extinction or
supercritical growth. Let $L_t^s$, $s\le t$, denote the branching
process with initial time $s$,  with $L_s^s=1$, which counts the number of
type-$1$'s at $t$ originating from a new type-$1$ at $s$. Then, the
total number of trait-$1$ species at $t$ is a random sum
\begin{equation}\label{poissonrepr}
L_t=\sum_{i: s_i\le t} L_t^{s_i},\quad t\ge 0,
\end{equation}
where $L_t^{s_i}$, $i\ge 1$, are independent copies of the type-$1$
branching process. Under cladogenetic splitting, the process $K_t$ is
independent of the number of $1$-species and (\ref{poissonrepr}) is a
Poisson sum representation of $L_t$.  The dynamics of an anagenetic
transition is more involved as $K_t$ decreases by one at each jump up
of $L_t$, and (\ref{poissonrepr}) is a self-regulating 
process rather than a Poisson process.  In both cases, however, the
expected number of $1$-species at $t$ is
\[
E_0(L_t)
=\delta\int_0^t E_0(K_s)E_1(L_t^s)\,ds ,\quad t\ge 0,
\]
where $E_0$ is the expectation starting from one species of trait-$0$
and $E_1$ is the expectation given an initial species of trait-$1$.
Similarly, 
\begin{equation} 
E_0(L_t|K_t>0) =\delta\int_0^t E_0(K_s|K_t>0)E_1(L_t^s)\,ds,
\label{expectedL}
\end{equation}
where, using (\ref{expectedK_s}),
\begin{equation}
E_0(K_s|K_t>0)= \frac{\lambda e^{\gamma_0 s}-\mu}{\gamma_0}
    + \frac{\lambda\mu(e^{\gamma_0 s}-1)(e^{\gamma_0(t-s)}-1)}
             {\gamma_0(\lambda e^{\gamma_0 t}-\mu)},
     \label{K_s}
\end{equation}
and
\begin{equation} 
E_1(L_t^s)=e^{\gamma_1(t-s)}.
\label{L_ts}
\end{equation}
Keeping the condition of at least one $0$-species at $t$, the expected
branch length of trait-$1$ species equals
\[
E_0(B_t|K_t>0) =\int_0^tE_0(\widetilde L_{s,t}|K_t>0)\,ds,
\]
where $\widetilde L_{s,t}$ is a summation of contributing reduced branching
processes $\widetilde L_{s,t}^u$, $u\le s\le t$, with $L^{u,t}_u=1$,
which originate from some point of the non-reduced, trait-$0$ tree
at time $u$.  Hence
\[
E_0(\widetilde L_{s,t}|K_t>0)=\delta \int_0^s E_0(K_u|K_t>0)
E_1(\widetilde L^u_{s,t})\,du.
\]
As in (\ref{condmeanreduced}), replacing $\lambda$, $\mu$, and $\gamma_0$
by $\lambda_1$, $\mu_1$,  and $\gamma_1$, respectively,
\begin{equation}
 \label{L_u}
E_1(\widetilde L^u_{s,t})
=\frac{1-p_0(t-u)}{1-v_{s-u,t-u}}
=\frac{\gamma_1e^{\gamma_1(t-u)}}{\lambda_1e^{\gamma_1(t-s)}-\mu_1}.
\end{equation}
Therefore
\begin{equation}
E_0(B_t|K_t>0)=\delta \int_0^t\int_0^s E_0(K_u|K_t>0)  E_1(\widetilde L^u_{s,t})\,duds,
 \label{branchlengthselfers}
\end{equation}
where
$E_0(K_u|K_t>0)$ is obtained in (\ref{K_s}), and $E_1(\widetilde L^u_{s,t})$ in (\ref{L_u}).

\subsection*{The Number of Trait-1 Clusters}\label{sec:cluster}

Let $C_t$ be the number of separate clusters of trait-$1$ species at
time $t$, that is, the number of sub-trees in $\mathcal T^1$ at $t$. Clearly, $1\le
C_t\le L_t$. The expected number of clusters is obtained by modifying
(\ref{expectedL}), as 
\begin{equation}
E_0(C_t|K_t>0)=\delta \int_0^t E_0(K_s|K_t>0) P_1(L^s_t>0)\,ds,
\label{nmbclusters}
\end{equation}
where $E_0(K_s|K_t>0)$ is given in (\ref{K_s}), and 
\begin{equation}\label{probL_t}
P_1(L^s_t>0)=\frac{\gamma_1}{\lambda_1-\mu_1 e^{-\gamma_1(t-s)}}.
\end{equation}
Now, by (\ref{L_ts}) and (\ref{probL_t}), 
it can be seen that for $0 \le \lambda_1 \le \mu_1$
\begin{equation}\nonumber
1 \leq  \frac{E_1(L_t^s)}{P_1(L^s_t>0)}  \leq 
\frac{\mu_1-\lambda_1 e^{-(\mu_1-\lambda_1)t}}{\mu_1-\lambda_1}.
\end{equation}
Hence, the ratio of the expected number of trait-$1$ 
species to the expected number of trait-$1$ clusters satisfies
\begin{equation}\label{L/C}
1\leq  \frac{E(L_t|K_t>0)}{E(C_t|K_t>0)}  \leq  
\frac{\mu_1-\lambda_1 e^{-(\mu_1-\lambda_1)t}}{\mu_1-\lambda_1},
\end{equation}
for each value of $p$ and $\delta$. It is straightforward to verify that 
\begin{equation}\label{cond:la1zero}
1=\frac{E(L_t|K_t>0)}{E(C_t|K_t>0)} \;\;\;\;\;\: \mbox{if and only if} \;\;\;\;\;\: \lambda_1=0.
\end{equation} 
A conclusion of (\ref{cond:la1zero}) is that if we observe a
phylogenetic tree where any species of trait-$1$ at $t$ forms its own
singleton cluster with no other trait-$1$ species as a closest neighbor 
(hence, $C_t=L_t$), it is natural to make the parameter estimation
$\lambda_1=0$.

\subsection*{Further Estimates for the Case when $\lambda_1=0$}

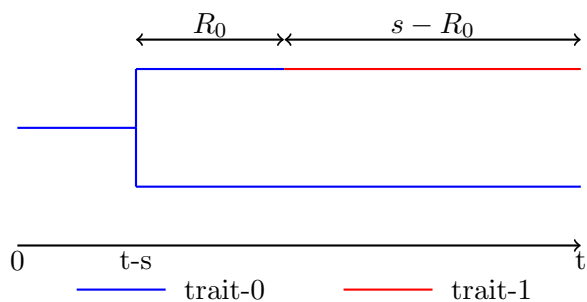
\begin{figure}[!t]
\centerline{
\begin{tikzpicture}[scale=0.39]
\draw [->] [thick](7,2) -- (26,2);
\node at (7,1.5) {0};
\node at (11,1.5) {t-s};
\node at (26,1.5) {t};
\node at (14,0.5) {trait-$0$};
\draw [blue][thick] (9,0.5) -- (12,0.5);
\node at (23,0.5) {trait-$1$};
\draw [red][thick] (18,0.5) -- (21,0.5);
\draw [blue][thick] (7,6) -- (11,6);
\draw [blue] [thick] (11,6)--(11,8);
\draw [blue] [thick] (11,8)--(16,8);
\draw [red] [thick] (16,8)--(26,8);
\draw [blue] [thick] (11,6)--(11,4);
\draw [blue] [thick] (11,4)--(26,4);
\draw [<->] [thick](11,9) -- (15.99,9);
\draw [<->] [thick](16.01,9) -- (26,9);
\node at (13.5,9.5) {$R_0$};
\node at (21,9.5) {$s-R_0$};
\end{tikzpicture}}
\caption{An observed pair of trait-$0$ (blue) and trait-$1$ (red) species at time $t$, 
with a common trait-$0$ ancestor at time $t-s$.} 
\label{fig:twospecies}
\end{figure}

We now examine the particular case, when each observed trait-$1$ species pairs up
with a species of trait-$0$, at the most recent branching bifurcation
point, hence the estimate $\lambda_1=0$. Let us consider such a pair
at time $t$, which traces back to a joint ancestor at time $t-s$. Since
$\lambda_1=0$, the joint ancestor is necessarily a species of trait-$0$. 
The total divergence time of the pair is $2s$. One branch of length $s$ is
trait-$0$ throughout, while the other branch divides into $s=R_0+(s-R_0)$, where
$R_0$ is the time spent as trait-$0$. In particular, if the splitting
event at $t-s$ produces one species of each trait, then $R_0=0$. 
This situation is illustrated in Figure \ref{fig:twospecies}. 

To evaluate correctly the differences between the two species due to
mutations, we need an estimate of the expected $R_0$. 
Suppose that we observe a total of $\ell$ species of trait-$1$ at
$t$, each having a trait-$0$ species as their nearest neighbor species backwards in the tree. 
Let the divergence times of each of the pairs be $t-s_i$,
$i=1,\dots,\ell$.  Let $R_0{(s_i)}$ be the corresponding times 
represented by trait-$0$ species since
divergence. Then 
\begin{equation}\label{estTself}
\begin{split}
A'_t&=A_t+\sum_{i=1}^\ell R_0{(s_i)},\\
 E(A'_t|K_t>0)&\approx E(A_t|K_t>0)+\sum_{i=1}^\ell r_0{(s_i)}, 
\end{split}
\end{equation}
where the $r_0{(s_i)}$'s denote the expected time spent as 
trait-$0$ on a branch of length $s$, and are computed as follows. 

Let $R$ denote an exponential random variable with rate $(1-p)\delta$,
and let $R_0$ have a mixed distribution so that $R_0=0$ with
probability $p\delta/(\lambda+p\delta)$ and $R_0$ is given by $R$
otherwise.  Also, let $V_0$ and $V_1$ be exponential extinction times
of rate $\mu_0$ and $\mu_1$, respectively.  Here, $R_0$ represents the
time as trait-$0$ in the branch ending up as trait-$1$, given that the
species survive to $t$.  Thus, using notation of the type
$E(X|A)=E(X,A)/P(A)$,

\begin{figure*}[!t]
\centerline{
\includegraphics[width=0.41\textwidth]{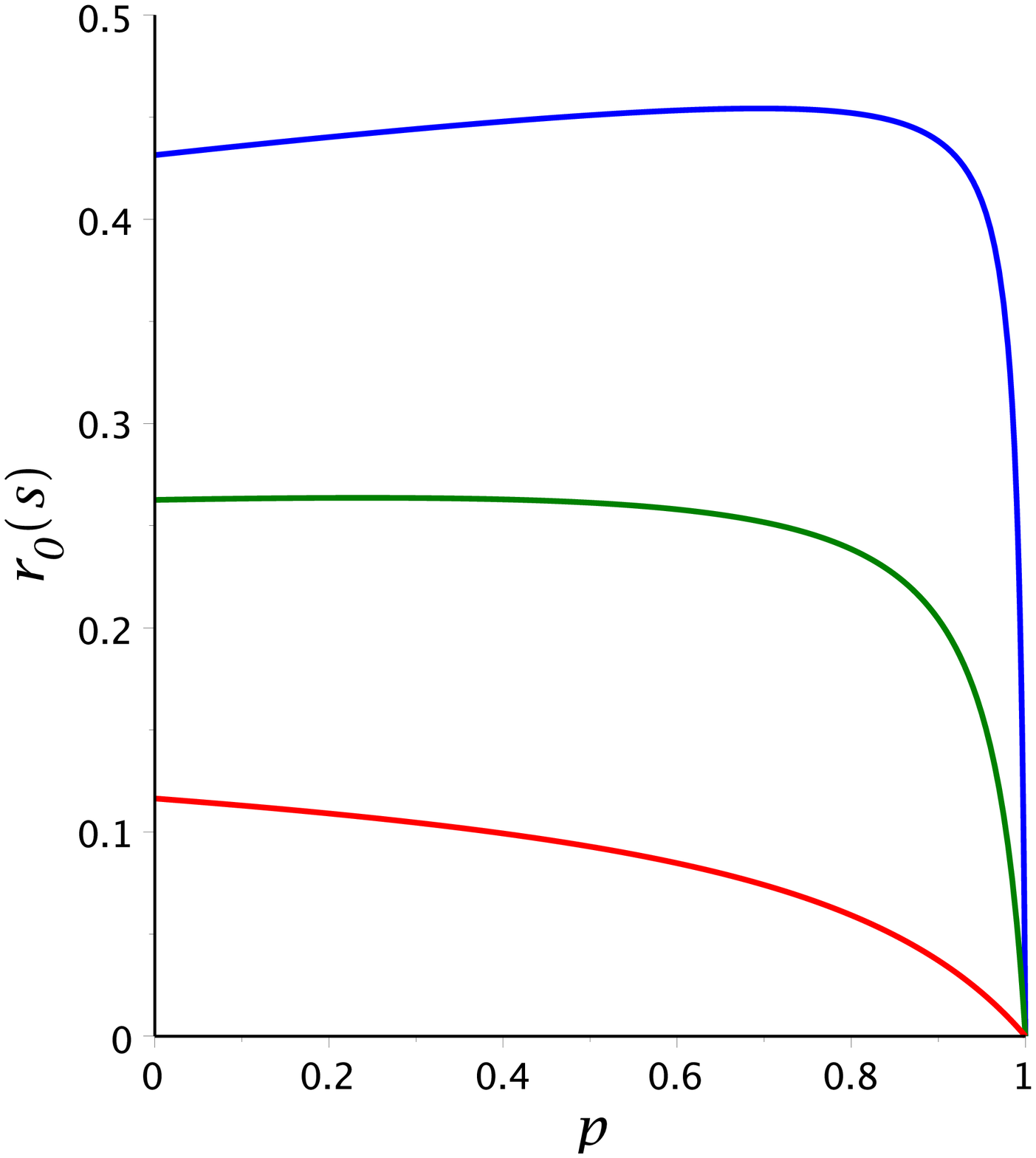}
\includegraphics[width=0.439\textwidth]{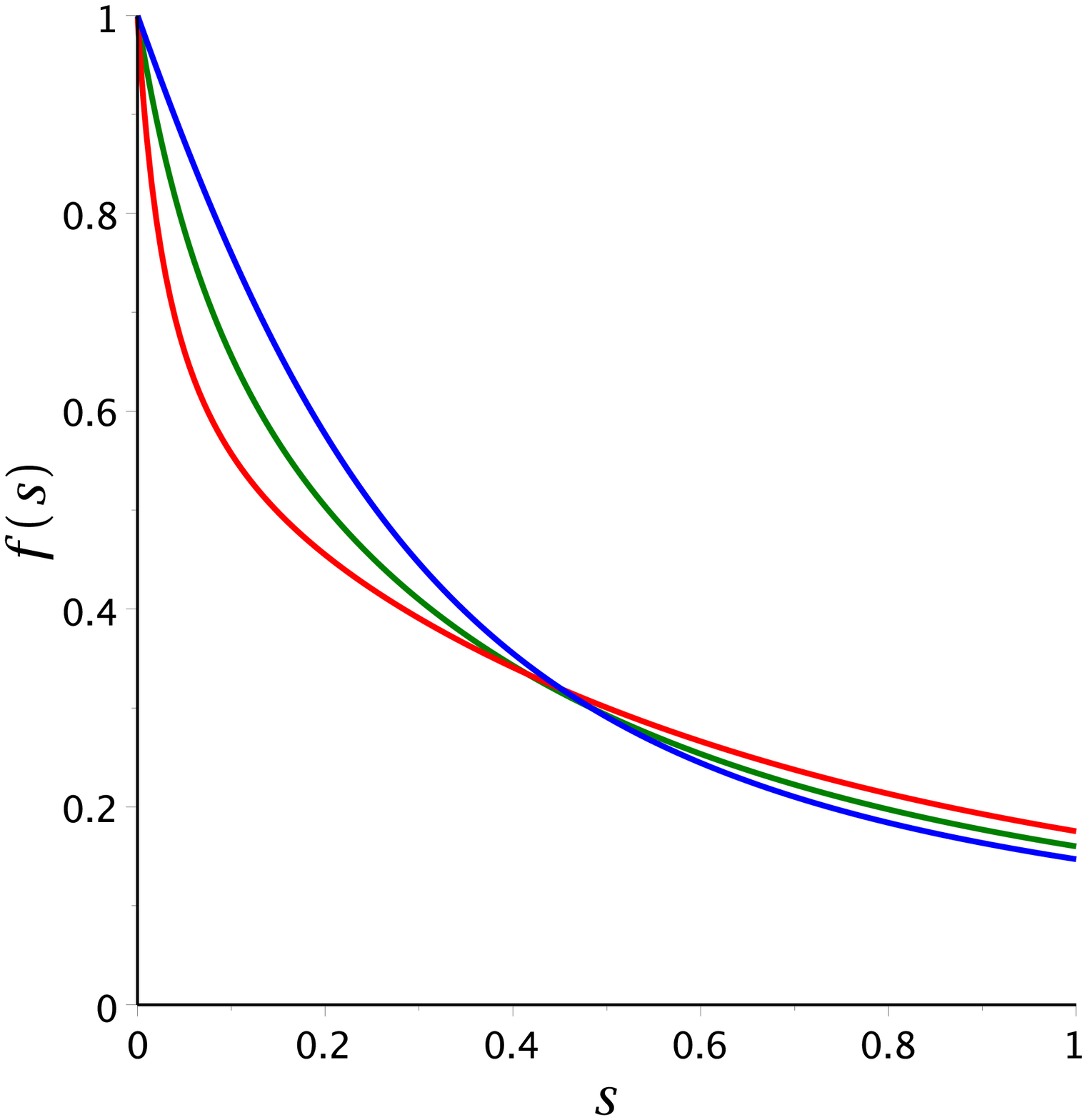}}
\caption{The left panel gives plots of $r_0(s)$ versus 
$p$ for $s=0.2$ in red, $s=0.4$ in green, and $s=0.6$ 
in blue, respectively. The right panel gives plot of 
$f(s)$ versus $s$ for $p=0.2$ in red, $p=0.4$ in green, 
and $p=0.6$ in blue, respectively. Here, $t=1$, 
$\omega_0=0.1$, $\omega_1=0.3$, $\delta=3$,  
$\lambda_0=8$, $\mu_0=4$, $\mu_1=\mu_0\omega_1/\omega_0=12$  
and $\lambda_1=0$.}
\label{fraction}  
\end{figure*}

\begin{align*} \nonumber
r_0(s)&=E(R_0|R_0<s, V_0>R_0, V_1>s-R_0)\\
  &= \frac{E(R_0,\,R_0<s, V_0>R_0, V_1>s-R_0)}{P(R_0<s, V_0>R_0, V_1>s-R_0)}.
\end{align*}

\noindent The expectation simplifies as

\begin{align*}
E&(R_0,\,R_0<s,V_0>R_0, V_1>s-R_0)\\
  &=\frac{\lambda}{\lambda+p\delta}\, E(R,\,R<s,V_0>R, V_1>s-R)\\
  &=\frac{\lambda}{\lambda+p\delta}\,\int_0^s r e^{-\mu_0 r}e^{-\mu_1(s-r)}(1-p)\delta e^{-(1-p)\delta r}\,dr.
\end{align*}

\noindent Furthermore,

\begin{align*}
P&(R_0<s, V_0>R_0, V_1>s-R_0)\\
&=\frac{p\delta}{\lambda+p\delta}P(V_1>s)+\frac{\lambda}{\lambda+p\delta}\,P(R<s, V_0>R, V_1>s-R)\\
&=\frac{p\delta e^{-\mu_1 s}}{\lambda+p\delta}+\frac{\lambda}{\lambda+p\delta}\int_0^s e^{-\mu_0 r-\mu_1(s-r)}
(1-p)\delta e^{-(1-p)\delta r}dr,
\end{align*}

\noindent and hence, recalling that $\mu=\mu_0+(1-p)\delta$, 
\[
r_0(s)=\frac{\lambda(1-p) \int_0^s r e^{(\mu_1-\mu)r} \,dr}
     {p + \lambda(1-p) \int_0^s e^{(\mu_1-\mu)r}\,dr},
\]
or, evaluating the integrals, 

\begin{align}
\label{r_0}
r_0(s)=\frac{\lambda(1-p)\Big((\mu_1-\mu)s - 1+ e^{-(\mu_1-\mu)s}\Big)}
     {p(\mu_1-\mu)^2e^{-(\mu_1-\mu)s} + \lambda(1-p)(\mu_1-\mu) (1-e^{-(\mu_1-\mu)s})}.
\end{align}

Figure \ref{fraction} (left panel) shows how $r_0(s)$ varies with $p$ 
for different values of $s$, when $\lambda_1=0$. In particular, 
it can be seen that when changes are not purely cladogenetic ($p<1$), 
a substantial fraction of the branch evolves as trait-$0$.

Define $f(s)$ to be the fraction of type-$1$, in an 
observed trait-$0$/trait-$1$ species pair, that is,  
\[
f(s)\coloneqq\frac{s-r_0(s)}{s}.
\]
Figure \ref{fraction} (right panel) gives an illustration of $f(s)$ versus 
$s$ for different values of $p$, with $\lambda_1=0$. It can be seen that $f(s)$ 
decreases with $s$ whenever $\mu_1>\mu=\mu_0+(1-p)\delta$. 
This means that proportionally, the longer the branch leading to type-$1$, 
the more recent the transition event would be.

\subsection*{Trait-0 Tree Conditioned on the number of Species}

Here, we compare the properties of the $0$-species tree derived
previously under the assumption of non-extinction at a fixed time,
$K_t>0$, to the corresponding properties assuming a fixed number of
trait-$0$ species at time $t$, $K_t=n$. This apart, the setting is the
same, and hence the single-type linear branching process $(K_t)_{t\ge 0}$,
$K_0=1$, representing the number of trait-$0$ species, has splitting rate
$\lambda=\lambda_0$ and extinction rate $\mu=\mu_0+(1-p)\delta$
restricted to the critical or supercritical case,
$\gamma_0=\lambda-\mu\ge 0$.  Let $t$ be fixed and condition on $K_t=n$.

Given $n$ trait-$0$ species at time $t$, the $n-1$ bifurcation
times, $S_1,\dots,S_{n-1}$, are the time intervals from the tips of
the tree at $t$ backwards until two species merge. The bifurcation
times of the full tree are the same as the bifurcation times of the
reduced tree, given $K_t=n$.  As an alternative, we may think of the
tree starting at the time of the most recent ancestor, and scale the
tree on the interval $[0,t]$.  In both cases, it turns out that the
joint distribution of the bifurcation times is the same as that of
$n-1$ i.i.d.\ observations sampled from a particular family of distribution
functions $F_t(s)$, $0\le s\le t$, depending on $\lambda$ and $\mu$
(\cite{thompson}; \cite{gernhard}). For the supercritical case, $\mu<\lambda$,
\begin{equation} \label{bifurcdistr_supercrit}
F_t(s)=\frac{\lambda-\mu e^{-\gamma_0t}}{1-e^{-\gamma_0t}}\,
\frac{1-e^{-\gamma_0 s}}{\lambda-\mu e^{-\gamma_0 s}},\quad
0\le s\le t,
\end{equation}
and for the critical case, $\mu=\lambda$,
\begin{equation} \label{bifurcdistr_crit}
F_t(s)=\frac{(1+\lambda) s}{t+\lambda s}, 
\quad f(t)=\frac{(1+\lambda)t}{(t+\lambda s)^2}, \quad 0\le s\le t.
\end{equation} 
Writing $S_{(k)}$, $1\le k\le n-1$, for the ordered bifurcation times
of the reduced type-$0$ tree and adding $S_{(0)}=0$ and
$S_{(n)}=t$, so that
\[
0=S_{(0)}\le S_{(1)}\le \dots \le S_{(n-1)}\le S_{(n)}=t,
\]
it follows that total branch length $A_t$ in (\ref{totalbranchout}) has the representation

\begin{align*} 
A_t&=n(S_{(1)}-S_{(0)})+(n-1)(S_{(2)}-S_{(1)})+\dots +1\cdot(S_{(n)}-S_{(n-1)})\\
   &=S_{(1)}+S_{(2)}+\dots+S_{(n-1)}+t\\
 &=S_{1}+S_{2}+\dots+S_{n-1}+t.
\end{align*} 

\noindent Hence
\begin{equation} \label{expectedA}
E(A_t|K_t=n)=(n-1)E[S]+t,
\end{equation}
where 
\begin{equation} \label{expectedS}
E(S)=\int_0^t(1-F_t(s))\,ds,
\end{equation}
 is obtained
from (\ref{bifurcdistr_supercrit}) or (\ref{bifurcdistr_crit}).
We may also use the {\it speciation times}
\[
T_k=t-S_{n-k},\quad T_{(k)}=t-S_{(n-k)},\quad k=0,\dots,n-1,
\]
as an alternative to bifurcation times.  The ordered speciation times
$T_{(k)}$, starting at $T_{(0)}=0$, are the successive branch time points of
the reduced tree $\widetilde K_{s,t}$ conditioned on $\widetilde
K_{t,t}=K_t=n$.  Here, all $T_k$, $1\le k\le n-1$, are independent and
identically distributed with distribution function

\begin{align*}
G_t(s)=P(T_k\le s)=P(S> t-s)=1-F_t(t-s)
=\frac{\gamma_0}{\lambda
  e^{\gamma_0(t-s)}-\mu} \frac{1-e^{-\gamma_0 s}}{1-e^{-\gamma_0 t}}.
\end{align*}
\noindent Also,
\begin{align}
E(\widetilde K_{s,t}|K_t=n)=1+E(\sum_{k=1}^{n-1} 1_{\{T_k\le s\}})=
 1+\sum_{k=1}^{n-1} P(T_k\le s)
   =1+(n-1)G_t(s)
    \label{fixed}
\end{align}
\noindent is consistent with (\ref{expectedA}) in the form
\begin{align} 
E(A_t|K_t=n)=\int_0^t (1+(n-1)G_t(s))\,ds
 =t+(n-1)\int_0^t 
   \frac{\gamma_0}{\lambda
  e^{\gamma_0(t-s)}-\mu} \frac{1-e^{-\gamma_0 s}}{1-e^{-\gamma_0
    t}}\,ds.
\label{Ebranchlengthgiven_n}
\end{align}

\subsection*{An Illustration Using Arbitrary Parameters}

\begin{figure*}[!t]
\centerline{
\includegraphics[width=.55\textwidth]{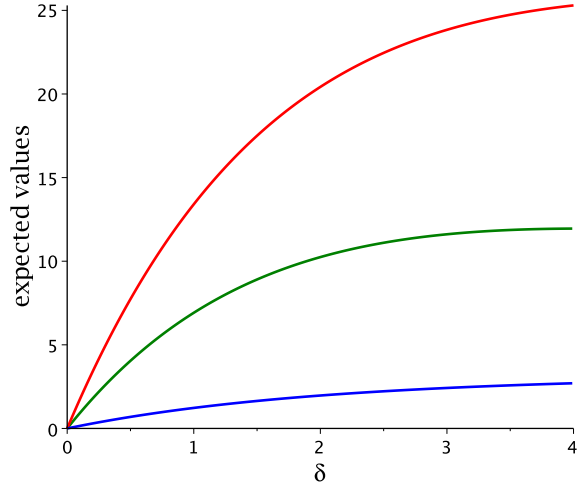}}
\caption{Plot of $E(L_t|K_t>0)$ versus $\delta$ in red; 
$E(C_t|K_t>0)$ versus $\delta$ in green; and $E(B_t|K_t>0)$ 
versus $\delta$ in blue. Here, $p=0.5$, $t=1$, $\omega_0=0.1$, 
$\omega_1=0.3$, $\lambda_0=8$, $\mu_0=4$, $\lambda_1=10$, $\mu_1=z\mu_0=12$ 
and $\mu=4+(1-p)\delta$. }
\label{fig:example}
\end{figure*}

To illustrate the species tree model, we first notice that the
trait-$0$ tree, corresponding to the blue colored sub-tree $\mathcal T^0$ in Figure
\ref{fig:treesplit}b, only depends on the parameters $\lambda$ and $\mu$.
In particular, the expected number of such species at $t$, given at
least one existing species, is given by (\ref{expectedK_s}) as 
\[
 E(K_t|K_t>0)=\frac{\lambda e^{(\lambda-\mu)t}-\mu}{\lambda-\mu}.
\]
Moreover, the expected branch length of the reduced tree,
corresponding to the blue-colored subtree $(\mathcal T^0)_t$ in Figure
\ref{fig:reducesplit}c, is obtained in (\ref{dndsweightout}), as 
\[
E(A_t|K_t>0)=\int_0^t \frac{\lambda e^{(\lambda-\mu) t}-\mu}{\lambda
  e^{(\lambda-\mu)(t-s)}-\mu}\,ds.   
\]
In case we are given $A_t=n$, then according to
(\ref{Ebranchlengthgiven_n}) the expected branch length equals
\[
E(A_t|K_t=n)=
t+\frac{n-1}{1-e^{-\gamma_0 t}} \int_0^t  \frac{\gamma_0
  (1-e^{-\gamma_0 s})} {\lambda e^{\gamma_0(t-s)}-\mu}\,ds.
\]
In order to comment on the functionals relating to the full tree, let
us take a fixed value of the probability $p$ of cladogenetic speciation
and assume that the mutation rates $\omega_0$ and $\omega_1$ are
known. Assuming now that point estimates of the parameters $\lambda$
and $\mu$ are known, it means that the trait-$0$ splitting rate
$\lambda_0=\lambda$ is known whereas the trait-$1$ splitting rate
$\mu_0=\mu-(1-p)\delta$ is given as a function of $\delta$. The model
assumption (\ref{c}) implies moreover that the ratio
\[
z=\frac{\mu_1}{\mu_0}=\frac{\omega_1}{\omega_0}
\]
is known, and therefore $\mu_1=z\mu-z(1-p)\delta$ is also a function
of $\delta$.  Various trait-$1$ functionals may now be studied at a fixed time
$t$ of observation as functions of $\delta$ and the remaining
parameter $\lambda_1$. For a set of arbitrary parameters $p$,
$\lambda_0$, $\mu_0$, $\omega_0$, $\omega_1$, and $\lambda_1$, Figure
\ref{fig:example} gives plots of the three functionals, which we will later use to
match with data, illustrating the dependence on the trait transition
intensity $\delta$.

\begin{figure*}[!t]
\centerline{
\includegraphics[width=0.458\textwidth]{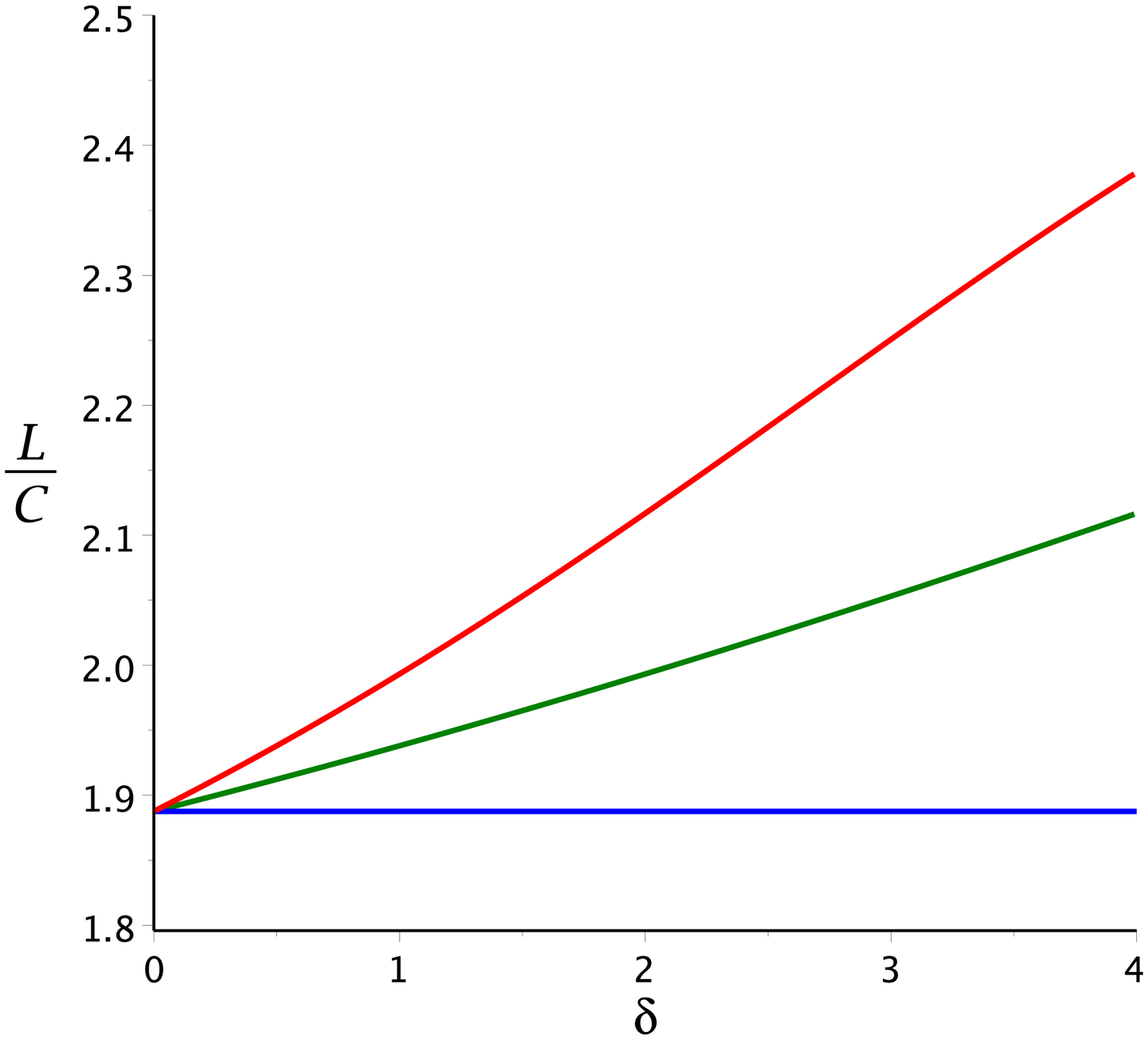}
\includegraphics[width=0.442\textwidth]{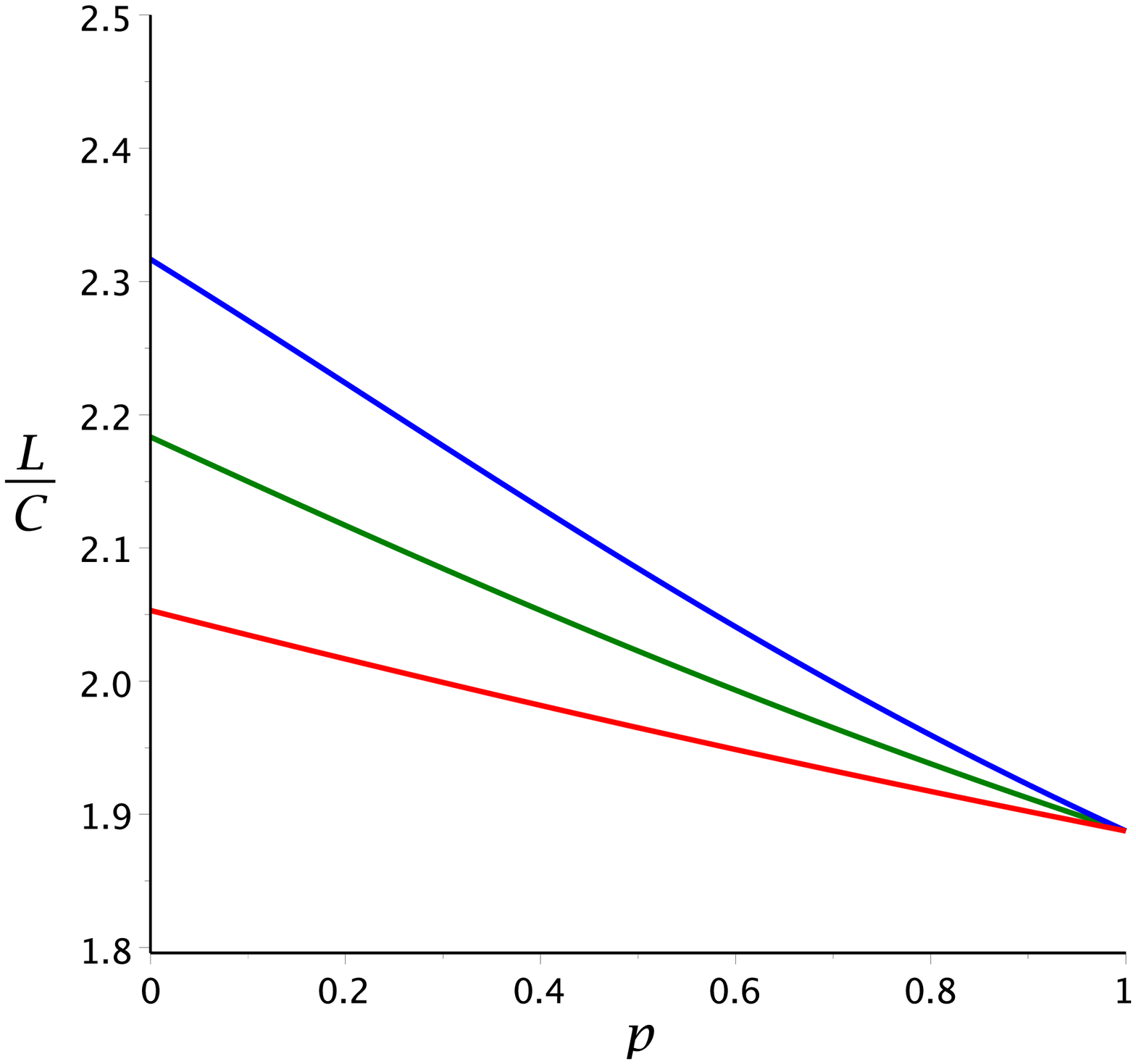}}
\caption{The left panel gives plots of the 
ratio $L/C=E(L_t|K_t>0)/E(C_t|K_t>0)$ versus 
$\delta$ for $p=0$ in red, $p=0.5$ in green, 
and $p=1$ in blue, respectively. The right 
panel  gives plots of the ratio $L/C$ versus 
$p$ for $\delta=1.5$ in red, $\delta=2.5$ in 
green, and $\delta=3.5$ in blue, respectively. 
Here, $t=1$, $\omega_0=0.1$, $\omega_1=0.3$, $\lambda_0=8$,
$\mu_0=4$ and $\lambda_1=10$.}
\label{fig2:example}
\end{figure*}

Using the same set of arbitrary parameters 
as in Figure \ref{fig:example}, we now obtain plots 
of the ratio  $E(L_t|K_t>0)/E(C_t|K_t>0)$ as a function 
of $p$ and $\delta$. For convenience, we use the shorthand 
notation $L/C$ to denote this ratio. 
Plots of $L/C$ as a function of $\delta$, for fixed values of $p$, are shown 
in left panel of Figure \ref{fig2:example}. Similarly, plots of 
$L/C$ versus $p$, for selected values of $\delta$, 
are given in right panel of Figure \ref{fig2:example}.
As shown in (\ref{L/C}), it can be seen that for any value of $p$ and $\delta$, 
the expected number of trait-$1$ species per cluster lies within a given bound. 
The upper bound is conservative, and for a wide range of arbitrary parameter 
values, $L/C$ remains closer to the lower bound,  
meaning that is is unlikely to observe large clades of trait-$1$ species.

 \section{Application to $dN/dS$}

In this section, for convenience, we refer to trait-$0$ as outcrosser
and trait-$1$ as selfer.
The $dN/dS$-ratio measures the normalized ratio of nonsynonymous to
synonymous substitutions. The total mutation intensity  splits in two
contributions, $\theta=\theta_\mathrm{syn}+\theta_\mathrm{non}$. The
precise fractions $\theta_\mathrm{syn}$, representing synonymous
mutations and $\theta_\mathrm{non}$, nonsynonymous mutations, can be
obtained from a detailed mutation model \cite{mugal_etal} or estimated from data.
  In the long run, synonymous substitutions build up
neutrally at scaled rate $\theta_\mathrm{syn}$.  The substitution rate
of nonsynonymous mutations, on the other hand, is reduced by sligthly
deleterious selection to $\omega_0\theta_\mathrm{non}$ for outcrossing
species and to $\omega_1\theta_\mathrm{non}$ for selfers,
$\omega_0<\omega_1<1$.  If we observe a single outcrossing species
known to exist over a fixed time duration $t$, the expected number of
substitutions are $\theta_\mathrm{non}\omega_0 t$ and
$\theta_\mathrm{syn} t$ for the two categories, and we understand the
normalized ratio to be simply $\omega_0$. Similarly, for a species
known to have been selfing over a fixed time interval, the
corresponding ratio is $\omega_1$.

\begin{figure*}[!t]
\centerline{\includegraphics[width=0.9\textwidth]{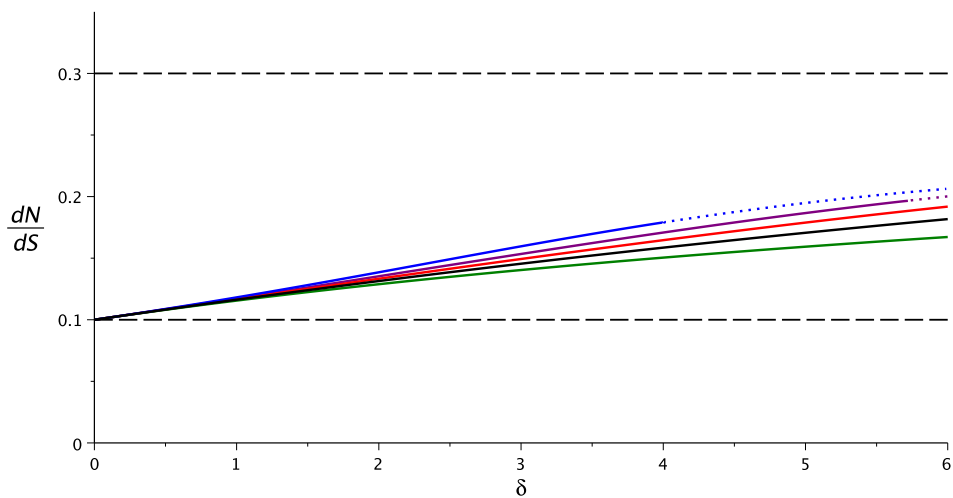}}
\caption{ Plots of $dN/dS$ versus $\delta$ for selected values of $p$. 
The blue curve is obtained for $p=0$; purple for $p=0.3$;
  red for $p=0.5$; black for $p=0.7$; and green for $p=1$.
  Here, $t=1$, $\omega_0=0.1$, $\omega_1=0.3$, $\lambda_0=8$,
  $\mu_0=4$ and $\lambda_1=10$. The dashed black lines represent 
  the value of $\omega_0$ and $\omega_1$. The dotted curves give those values 
  of $dN/dS$ which are obtained for a sub critical process (that is, when 
  $\gamma_0<0$).}
  \label{dNdSvsdelta_p2}
\end{figure*}

To capture in more detail, the accumulation of fixed mutations in the
species tree, we run independent Poisson processes along the
branches. First, a collection of Poisson points $\mathcal{S}$ with
intensity $\theta_\mathrm{syn}$ placed on top of the entire tree,
represents synonymous substitutions.  Next, nonsynonymous substitutions
are generated by a Poisson measure $\mathcal{N}^{(0)}$
with intensity $\omega_0\theta_\mathrm{non}$ along all
outcrossing branches, and by a measure $\mathcal{N}^{(1)}$ with intensity
$\omega_1\theta_\mathrm{non}$ along the branches representing selfers.  
The random variable $\mathcal{S}_{t}=\mathcal{S}(\mathcal T_t)$ counts the number of synonymous
substitutions in the species tree reduced at $t$. Thus, 
\begin{align*}
\mathcal{S}_{t}&=\mbox{$\#$ accumulated synonymous substitutions in all existing species at
  $t$}, \quad t\ge 0,
\end{align*}
is a Poisson process modulated by the stochastic intensity
$\theta_\mathrm{syn}\, (A'_t+B_t)$. Indeed, conditionally, given $t$
and given the total branch lengths $A'_t$ and $B_t$ of outcrossers and selfers
in the reduced tree, introduced in (\ref{totalbranchoutplus}) and
(\ref{totalbranchself}), $\mathcal{S}_{t}$ has a Poisson distribution with mean
$\theta_\mathrm{syn}(A'_t+B_t)$.  Similarly,

\begin{align*}
\mathcal{N}^{(i)}_t&=\mbox{$\#$ nonsynonymous substitutions of type-$i$ up to
  time $t$}, \quad i=0,1,
\end{align*}

\noindent are Poisson processes modulated by the random intensities
$\theta_\mathrm{non}\omega_0\, A'_t$ and $\theta_\mathrm{non}\omega_1\,
B_t$, respectively.  

\begin{figure*}[!t]
\centerline{\includegraphics[width=0.9\textwidth]{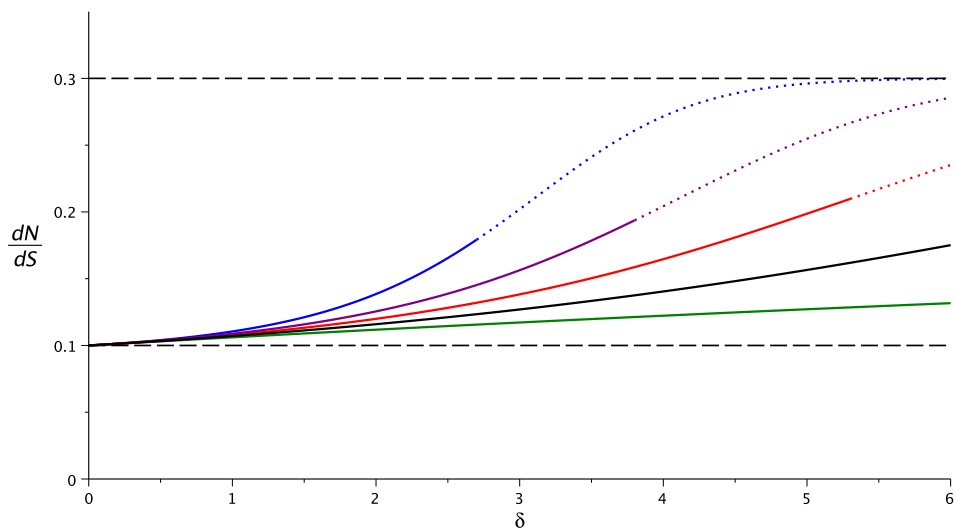}}  
\caption{ Plots of $dN/dS$ versus $\delta$ for selected values of $p$. 
The blue curve is obtained for $p=0$; purple for $p=0.3$;
  red for $p=0.5$; black for $p=0.7$; and green for $p=1$.
  Here, $t=1$, $\omega_0=0.1$, $\omega_1=0.3$, $\lambda=8$,
  $\mu=6$ and $\lambda_1=10$. The dashed black lines represent 
  the value of $\omega_0$ and $\omega_1$. The dotted curves are those values 
  of $dN/dS$ which are not compatible with the biological assumptions (that is, when 
  $r_1>0$). In this example, the trait-$0$ tree parameters, $\lambda$ and $\mu$, 
  are fixed, and these give $E(A_1|K_1>0)\approx 5.7$. Hence, the outcrosser tree is preserved. }
  \label{dNdSvsdelta_p}    
\end{figure*}

In our context, it is natural to associate $dN/dS$ with the average
number of substitutions, which have occurred anywhere in the species
family tree and are observable today.  Substitutions observable
today must have occurred on the reduced tree.  Of course, a meaningful
$dN/dS$ concept is naturally conditioned on survival of some species
today. This leads us to considering the $dN/dS$-ratio

\begin{align*}
dN/dS
&\sim\frac{(E(\mathcal{N}_t^{(0)}|K_t>0)+E(\mathcal{N}_t^{(1)}|K_t>0)/\theta_\mathrm{non}}
{E(\mathcal{S}_t|K_t>0)/\theta_\mathrm{syn}}\nonumber\\
&=\frac{\omega_0E(A'_t|K_t>0)+\omega_1E(B_t|K_t>0)}
{E(A'_t+B_t|K_t>0)}. 
\end{align*}

\noindent The conditioning scheme of assuming at least one outcrosser at $t$,
$K_t>0$, is to some degree, arbitrary. Alternatives, such as assuming
$K_t=k$ or imposing a condition involving both $K_t$ and $L_t$, are
equally natural. Our choice is computationally more convenient and
hence we define
\[ 
dN/dS|_t=\frac{\omega_0E(A'_t|K_t>0)+\omega_1E(B_t|K_t>0)}
{E(A'_t|K_t>0)+E(B_t|K_t>0)}.
\] 
Typically it is straightforward to estimate the total branch length in the
denominator, given by  
\[
T_\mathrm{tot}=E(A'_t|K_t>0)+E(B_t|K_t>0),
\]
and also the ratio
\begin{equation}
q^0_t=\frac{E(A_t|K_t>0)}{E(A'_t|K_t>0)+E(B_t|K_t>0)},
\label{ratio}
\end{equation}
while the desired ratio of expected values is
\begin{equation}
q_t=\frac{E(A'_t|K_t>0)}{E(A'_t|K_t>0)+E(B_t|K_t>0)}\ge q^0_t,
\label{ratio2}
\end{equation}
for which 
\[
dN/dS|_t=q_t \omega_0+(1-q_t)\omega_1.
\]
The value of $q_t$ in (\ref{ratio2}) may be be computed
when $\lambda_1=0$, since in this case, $E(A'_t|K_t>0)$ can be 
obtained using (\ref{estTself}). 

For $t=1$, fixed mutation rates $\omega_0$ and $\omega_1$,
and arbitrary values of parameters $\lambda_0$, $\mu_0$, and 
$\lambda_1$, Figure \ref{dNdSvsdelta_p2} illustrates the shape of 
$dN/dS$ versus $\delta$, for selected values of $p$. Here, 
$\mu_1=\mu_0\omega_1/\omega_0$ is fixed, whereas 
$\mu$, obtained by using the relation $\mu=\mu_0+(1-p)\delta$, 
varies with $p$ and $\delta$.  
Figure \ref{dNdSvsdelta_p} shows similar types of plots, 
obtained by using the same parameters as in Figure \ref{dNdSvsdelta_p2}, 
except that here, $\mu$ is fixed, while 
$\mu_0=\mu-(1-p)\delta$ and $\mu_1=\mu_0\omega_1/\omega_0$ 
are allowed to vary with $p$ and $\delta$. Since the values of 
$\lambda$ and $\mu$ are fixed in Figure \ref{dNdSvsdelta_p}, 
the outcrosser tree remains preserved in this case.

\section{ Relating Model to Species Tree Data}

This section discusses how to relate our theoretical study 
of random species trees to given data sets from 
two plant families, Geraniaceae and Solanaceae, in order to provide
a first test of the relevance of the mathematical modeling.   
In each case, a sequence data set is available for an observed
selfer-outcrosser species family of total size $m$, consisting of $k$
outcrossing and $l$ selfing species, such that $k+l=m$. 
The data set for the Geraniaceae family has a known outgroup that helps
placing the origin of the family. For the Solanaceae data set, the outgroup
information is missing, and hence we insert a virtual root for the
tree.

\subsection*{Geraniaceae Family}

\begin{figure}[!t]
\centerline{\includegraphics[width=0.7\textwidth]{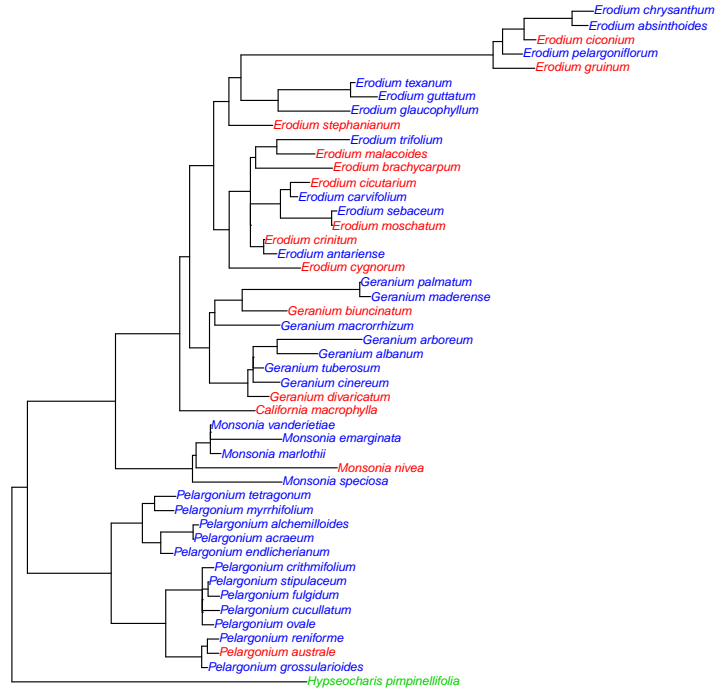}}
\caption{  Phylogenetic tree consisting of $47$ species in the
  Geraniaceae family and the outgroup. Outcrossers are colored in blue, selfers in
  red, and the outgroup in green.
  }
\label{big_rooted}
\end{figure}

The sequence data set for the Geraniaceae family, 
obtained from \cite{glemin_muyle}, 
consists of $1425$ codons in an interleaved format, 
a known outgroup, and $m=47$ species, 
of which $k=33$ are outcrossers and $\ell=14$ are
selfers.

\paragraph{\textit{1. Estimating the tree characteristics.}---}
 The phylogeny analysis software PhyML \cite{guindon_gascuel}, 
 with nucleotide substitution model GTR, branch support set to aLRT, 
 BIONJ starting tree, and tree searching operation NNI, was used to 
 construct the phylogenetic tree, which is given in Figure \ref{big_rooted}. 
The program PAML \cite{yang} was used to obtain $Q=0.0955$, 
defined as the estimate of the global $dN/dS$ ratio over the whole tree 
(excluding the outgroup). The outcrosser species sub-tree, was then obtained
by removing all selfing branches from the initial tree in Figure \ref{big_rooted}. 
PAML was used on this outcrosser sub-tree and the corresponding 
sequence data, which yielded the $dN/dS$-value, 
essentially an estimate of $\omega_0$, as $\omega_0=0.075$. 
The observed ordering $\omega_0<Q$ is consistent 
with our basic hypothesis that $\omega_0\le \omega_1$.

\begin{figure*}[t!]
\centerline{\includegraphics[width=0.85\textwidth]{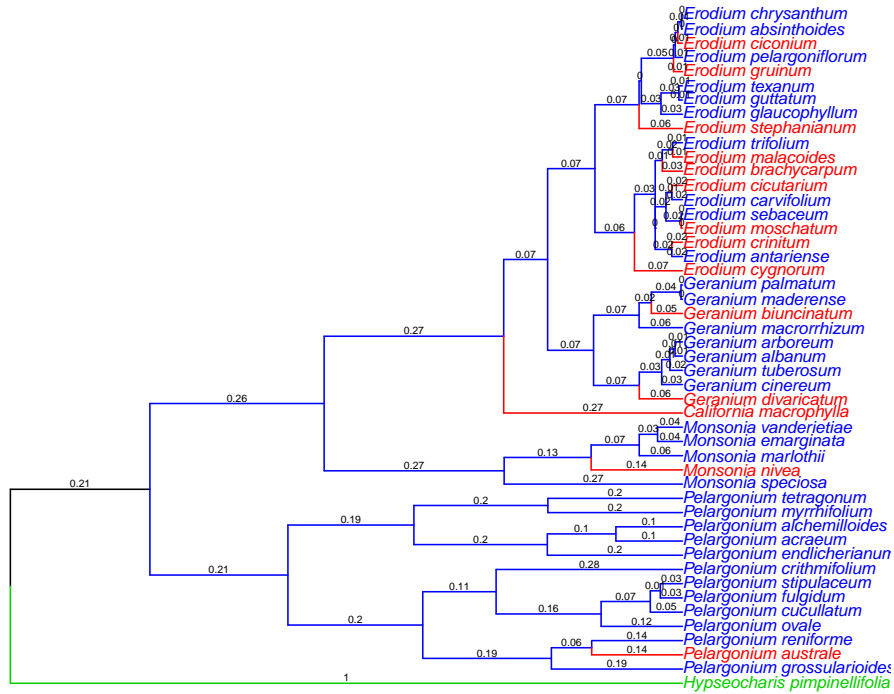}} 
\caption{Ultrametric phylogenetic tree consisting of $47$ species in the Geraniaceae family and the outgroup. Outcrosser 
branches are colored blue; the selfer branches, from bifurcation time onward,
are given in red; the outgroup branch is colored in green, and the
`virtual' outcrosser branch in black. The numbers on each branch represent the branch length.}
\label{big_rootedultra}  
\end{figure*}

\begin{figure*}[!t]
\centerline{\includegraphics[width=0.8\textwidth]{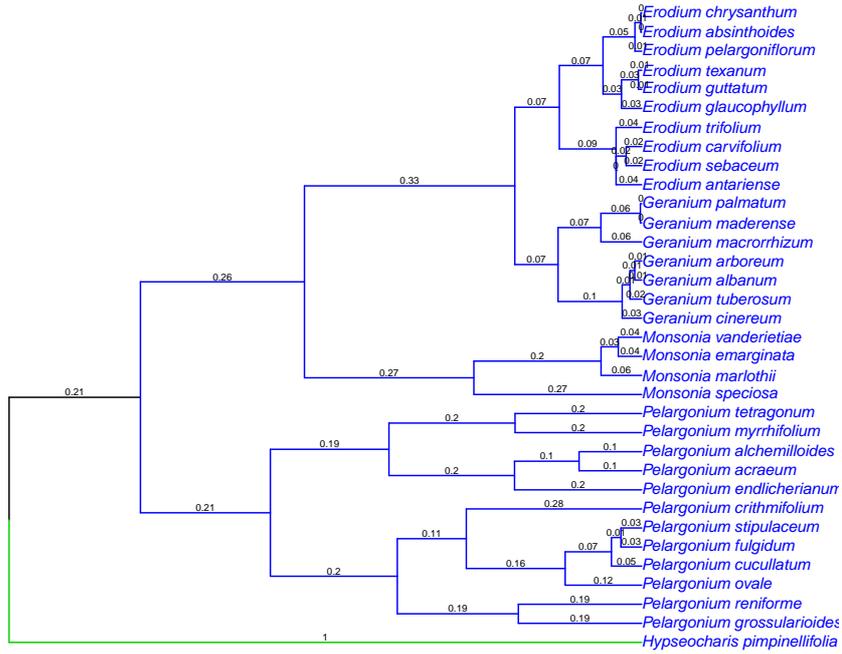}} 
\caption{Ultrametric outcrosser tree consisting of $33$ species 
in the Geraniaceae family and the outgroup. Outcrosser branches 
are in blue, the outgroup branch in green, and the virtual outcrosser branch is colored black. 
The number given on each branch represents the branch's length.}
\label{small_rootedultra} 
\end{figure*}

The tree in Figure \ref{big_rooted} was then made ultrametric with 
the function `chronos' of the package `ape' 
\cite{paradis_etal} in the software R \cite{R}, 
and normalized to have $t=1$. This ultrametric tree is given in
Figure \ref{big_rootedultra}. The length of the branch, leading 
from the root of the tree to the first speciation time, was estimated 
using the outgroup species, and can be considered a `virtual' branch length, 
used to match the mathematical model to our phylogenetic analysis. 
The total branch length of the tree, excluding the outgroup, is $T_\mathrm{tot}=6.926$. 
By removing selfing branches from the ultrametric tree in Figure
\ref{big_rootedultra}, the ultrametric version of the outcrossing
sub-tree at time $t$ was also obtained, as given in Figure
\ref{small_rootedultra}. From this outcrossing sub-tree, 
corresponding to $(\mathcal T^0)_t$, we record: (a)
the observed ordered bifurcation times $0\le s_{(1)}\le\dots \le
s_{(n-1)}\le s_{(n)}=t$, given in Fig. \ref{small_rootedultra}; 
(b) the observed total branch-length 
\[
T^{(0)}_\mathrm{tot}=s_{(1)}+\dots+s_{(n)}=6.045; 
\] 
and, (c) the sample mean of the bifurcation times
\[
\bar s=(s_{(1)}+\dots +s_{(n-1)})/(n-1)=0.1577.
\] 

\paragraph{\textit{2. Obtaining the lower bound for $\omega_1$.}---}
The quotient
\[
q^0=T^{(0)}_\mathrm{tot}/T_\mathrm{tot}=0.873
\]
is a numerical estimate of the ratio in (\ref{ratio}), interpreted as the
minimal fraction of outcrossing branches in the full species tree.  
Since $Q$ is a point estimate of $dN/dS|_1$, we have
\begin{equation}\label{Q}
Q=q \omega_0+(1-q)\omega_1,\quad q\ge q^0
\end{equation}
where $q$ is the actual fraction of outcrossing branches in the full
tree. In view of (\ref{Q}), with $Q$ and  $\omega_0$
both known,  $\omega_1$ is an increasing function of $q$, and hence
we have the lower bound
\begin{equation}  \label{om1bound}
\omega_1=\frac{Q-\omega_0 q}{1-q}\ge 
\frac{Q-\omega_0 q^0}{1-q^0}= 0.236,\quad q\ge q^0.
\end{equation}

\begin{figure}[!t]
\centerline{\includegraphics[width=0.5\textwidth]{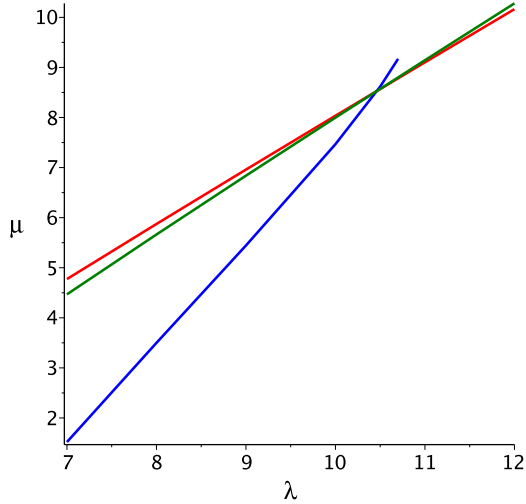}}
\caption{Numerical solutions for method i) plotted in green, for
  method ii) in blue, and for method iii) in red. The three lines approximately meet at 
  a point where $\lambda=10.462$ and $\mu=8.526$.}
\label{lambdamu}    
\end{figure}

\begin{figure*}[!t]
\centerline{
\includegraphics[width=0.46\textwidth]{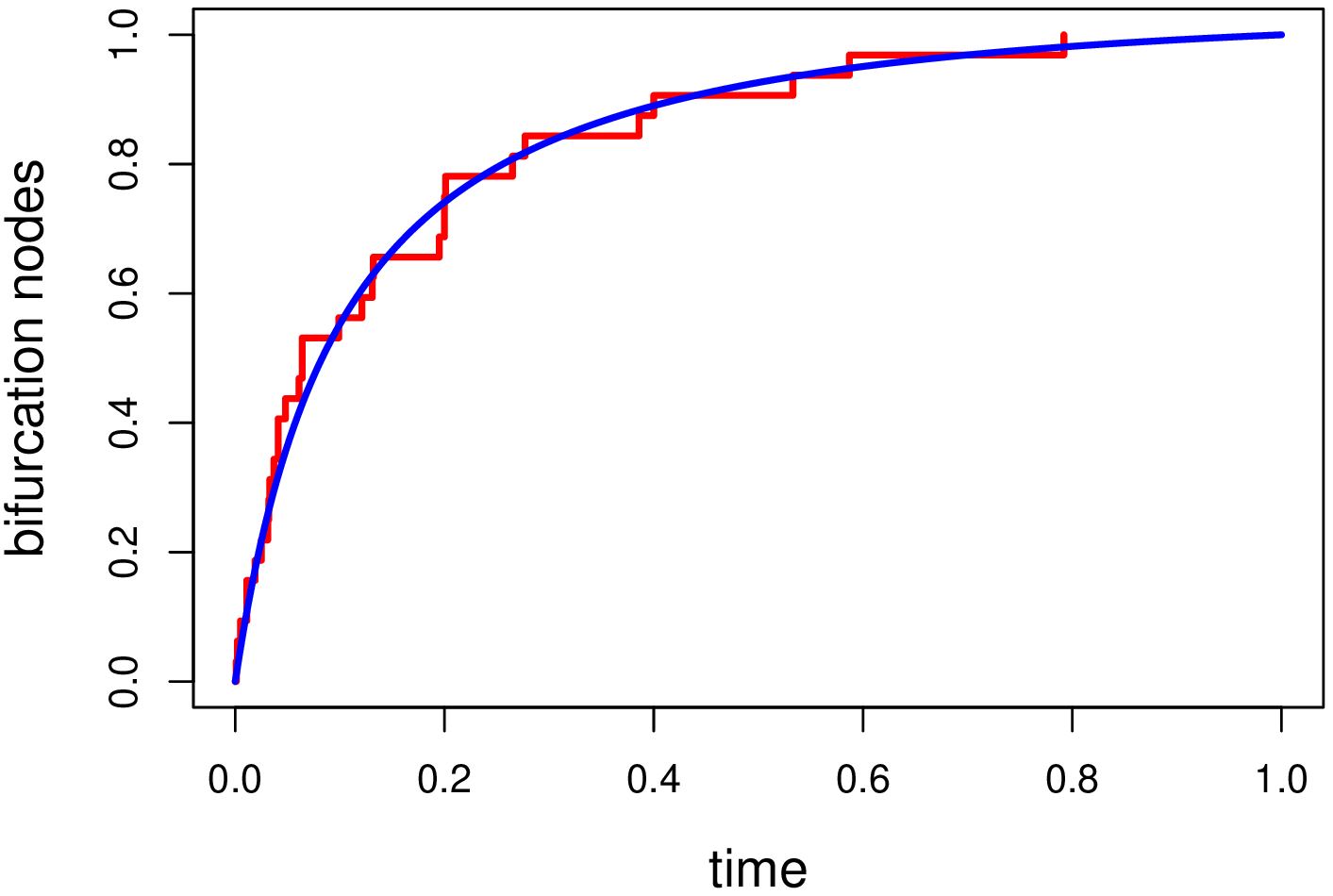}
\includegraphics[width=0.46\textwidth]{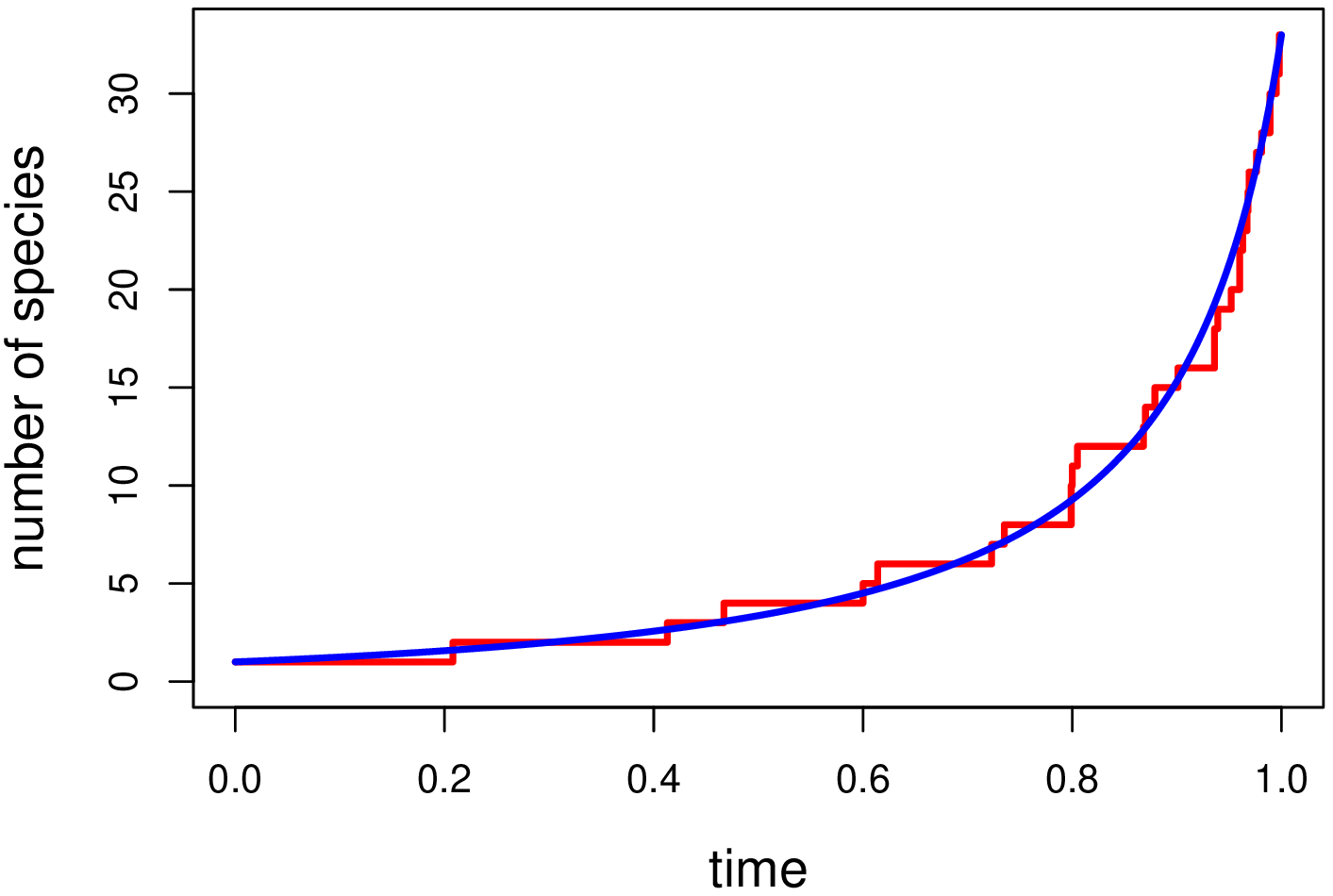}}
\caption{ In the left panel, the red plot gives the empirical distribution of the bifurcation times
  $(s_1,\dots,s_{32})$ of the tree given in Figure \ref{small_rootedultra} and the blue graph is a plot of 
  (\ref{bifurcdistr_supercrit}) with $t=1$, $\lambda=10.462$ and
  $\mu=8.526$. In the right  panel, the red graph is a plot of the number of species versus time for  
the tree given in Figure \ref{small_rootedultra}
and the blue graph is a plot of (\ref{fixed}) with $t=1$, $\lambda=10.462$ and $\mu=8.526$. }
\label{bifur}
\end{figure*}

\paragraph{\textit{3. Estimating the outcrosser parameters, $\lambda$ and $\mu$.}---}
We list three methods for extracting admissible pairs
$(\lambda,\mu)$ consistent with data. By combining
these methods, the parameter space is further reduced, 
leading to reasonable estimates of the separate parameters $\lambda$
and $\mu$. The methods are

\noindent
i) Use $k=33$ as a point estimate of $E(K_t|K_t>0)$ at $t=1$. By (\ref{expectedK_s}), this
gives 
\begin{equation}\label{outnumber}
\frac{\lambda e^{(\lambda-\mu)}-\mu}{\lambda-\mu}=33.
\end{equation}
ii)  Apply the relation (\ref{expectedS}) using $\bar s=0.1577$ as an estimate
of $E(S)$, that is,  estimate $(\lambda,\mu)$ from
\begin{align}
\label{bifurctime}
\int_0^1(1-F_1(s))\,ds =\int_0^1 \big(1-\frac{\lambda-\mu
  e^{-\gamma_0}}{1-e^{-\gamma_0}}\,\frac{1-e^{-\gamma_0
    s}}{\lambda-\mu e^{-\gamma_0 s}}\Big)\,ds = 0.1577. 
\end{align}
iii) Apply the branch length statistics $T^{(0)}_\mathrm{tot}=6.045$ as an
  approximation for $E(A_t|K_t>0)$ at $t=1$ obtained in (\ref{dndsweightout}), 
and hence solve for  $(\lambda,\mu)$ as
\begin{equation}\label{branchestimation}
\int_0^1 \frac{\lambda e^{\gamma_0}-\mu}{\lambda
  e^{\gamma_0s}-\mu}\,ds=6.045. 
\end{equation}
The $(\lambda,\mu)$ pairs obtained numerically in each of
the three cases are plotted in Figure \ref{lambdamu}.
By comparing the three estimation methods, 
we obtain the parameter estimates $\lambda\approx 10.462$ and
$\mu\approx 8.526$. Figure \ref{bifur} (left panel) shows that 
the empirical distribution of the bifurcation times $(s_1,\dots,s_{32})$ of the
ultrametric outcrosser tree, fits rather well the distribution
function $F_1(s)$, $0\le s\le 1$, of the supercritical
birth and death model in (\ref{bifurcdistr_supercrit}), when
$\lambda=10.462$ and $\mu=8.526$. Similarly,  
Figure \ref{bifur} (right panel) gives a plot of the number of species 
in the ultrametric outcrosser tree, along with a plot 
of the expected values in (\ref{fixed}), when $\lambda=10.462$  
and $\mu=8.526$.

\paragraph{\textit{4. Estimating the selfer species rates.}---} 
By the definition of $q$, and using (\ref{Q}), the expected selfer
branch length equals
\begin{equation}\label{B_relation}
E(B_1|K_1>0)=(1-q)T_\mathrm{tot}=\frac{Q-\omega_0}{\omega_1-\omega_0}T_\mathrm{tot}.
\end{equation}
Also, by (\ref{c}), and using the rate of removal 
of outcrossers ($\mu=\mu_0 + (1-p)\delta$), 
the exinction rate, $\mu_1$, of
selfers is a function of $p$, $\delta$, and $\omega_1$,  
\begin{equation}\label{mu1}
\mu_1=\frac{\omega_1}{\omega_0}\,\mu_0=\frac{\omega_1}{\omega_0}(\mu-(1-p)\delta). 
\end{equation} 
The Geraniaceae data set has the particular feature that
the number of selfing species coincides with the number of selfing
clusters, both $14$. In other words, each observed selfer, at
$t=1$, is joined to an outcrosser, at
the most recent bifucation point. In
particular, using (\ref{cond:la1zero}), we infer the estimate
$\lambda_1=0$. 

Now, by rewriting (\ref{L_u}) and (\ref{branchlengthselfers}) for $t=1$ and 
$\lambda_1=0$, we obtain  
\begin{equation}
E(B_1|K_1>0)=\delta\int_0^1\int_0^s E(K_u|K_1>0) e^{-\mu_1(1-u)}duds.
 \label{expectedlengthselfer}
\end{equation}
The right hand side of (\ref{expectedlengthselfer}) equated to the right hand side of
(\ref{B_relation}) gives
\begin{equation}
\delta \int_0^1\int_0^s E(K_u|K_1>0)e^{-\mu_1(1-u)}duds
=\frac{Q-\omega_0}{\omega_1-\omega_0}T_\mathrm{tot},
\label{Bcheck}
\end{equation}
where $E(K_u|K_1>0)$ is obtained in (\ref{K_s}). 

Similarly, (\ref{expectedL}) and (\ref{L_ts}) can be rewritten for $t=1$ and 
$\lambda_1=0$, as
\begin{equation} 
E(L_1|K_1>0) =\delta\int_0^1 E(K_s|K_1>0)e^{-\mu_1(1-s)}\,ds.
\label{expectedL1}
\end{equation}
Making the identification $E(L_1|K_1>0)=\ell=14$, we obtain from 
the right hand side of (\ref{expectedL1})
\begin{equation}\label{ellcheck}
\delta\int_0^1 E(K_s|K_1>0)e^{-\mu_1(1-s)}\,ds=14,
\end{equation}
where $E(K_s|K_1>0)$ is obtained in (\ref{K_s}), as before.

Furthermore, by invoking relation
(\ref{estTself}), valid for $\lambda_1=0$, we have the estimate $\hat
q$ of $q$, given by 
\begin{equation}\nonumber
\hat q=q^0+\frac{1}{T_\mathrm{tot}}\sum_{i=1}^\ell r_0{(s_i)},
\end{equation} 
where the selfer bifurcation times, $s_i=(s_1,\dots,s_{14})$, are shown in
Figure \ref{big_rootedultra} and $r_0{(s_i)}$, $i=1\dots14$, are obtained in (\ref{r_0}). Here, $\sum_i s_i=0.881$
whereas $T^{(1)}_\mathrm{tot}=0.881-\sum_i r_0{(s_i)}$ is the total expected branch length
represented by selfers. Now, each $(p,\delta,\omega_1)$ gives a
$\hat q$. On the other hand, each $\hat q$ yields an $\omega_1$, using 
\begin{equation}\label{om1check}
\hat \omega_1=\frac{Q-\hat q \omega_0}{1-\hat q}.
\end{equation}

\paragraph{\textit{5. Estimating the parameters $p$, $\delta$ and $\omega_1$.}---}

\begin{figure}[!t]
\centerline{\includegraphics[width=0.7\textwidth]{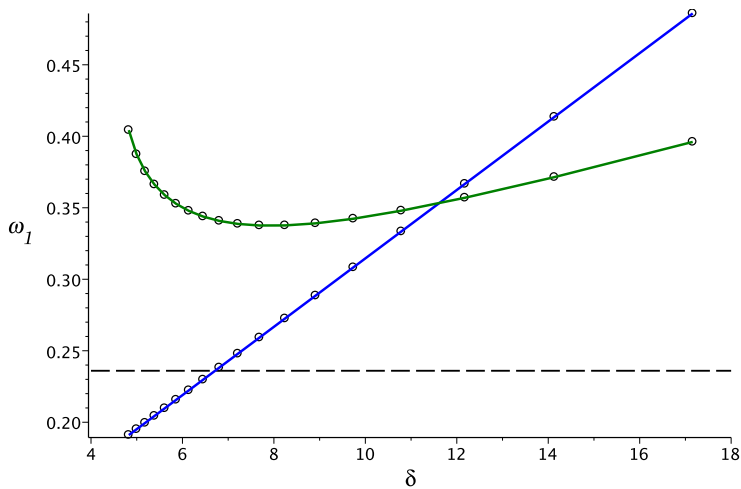}}
\caption{Plot of $\omega_1$ versus $\delta$ with fixed $p$-values. The blue 
graph represents pairs $(\omega_1,\delta)$ satisfying (\ref{Bcheck}) and
(\ref{ellcheck}). The green graph represents $(\omega_1,\delta)$ pairs which satisfy 
(\ref{om1check}). The circular marks on both graphs are successive $p$-values $0,
0.05,0.10,\dots,0.85$. The dashed line marks the lower bound for $\omega_1$. }
\label{fig:gerparest}
\end{figure}

For any fixed $p$, we are now in position to find admissible pairs
$(\omega_1,\delta)$ which satisfy the two equations (\ref{Bcheck}) and
(\ref{ellcheck}). These pairs are plotted as the blue line in Figure
\ref{fig:gerparest}, with successive $p$-values indicated as circles along the line starting
from the lower left. It is seen that all solutions with $p> 0.39$
are consistent with the bound $\omega_1\ge 0.236$ obtained in
(\ref{om1bound}).  By going through the $p$-values once more, however,
and evaluating $\hat\omega_1$ according to (\ref{om1check}), one
obtains the green line in Figure \ref{fig:gerparest}. At the
crossing point of the blue and green curves, a unique combination 
of parameters, which satisfy all criteria set up in this analysis, are obtained. 
These are $p=0.732$, $\omega_1=0.353$, and $\delta=11.623$.

\paragraph{\textit{6. Conclusions}---}As a final result, we obtain the following
branching tree rates 

\[
\lambda_0 =\lambda=10.462,\quad \lambda_1=0, \quad
 \mu_0=5.411, \quad \mu=8.526,  \quad \mu_1=25.468,
 \]
 \[ 
\delta=11.623,\quad  p=0.732.
\]

\begin{figure*}[!t]
\centerline{
\includegraphics[width=0.45\textwidth]{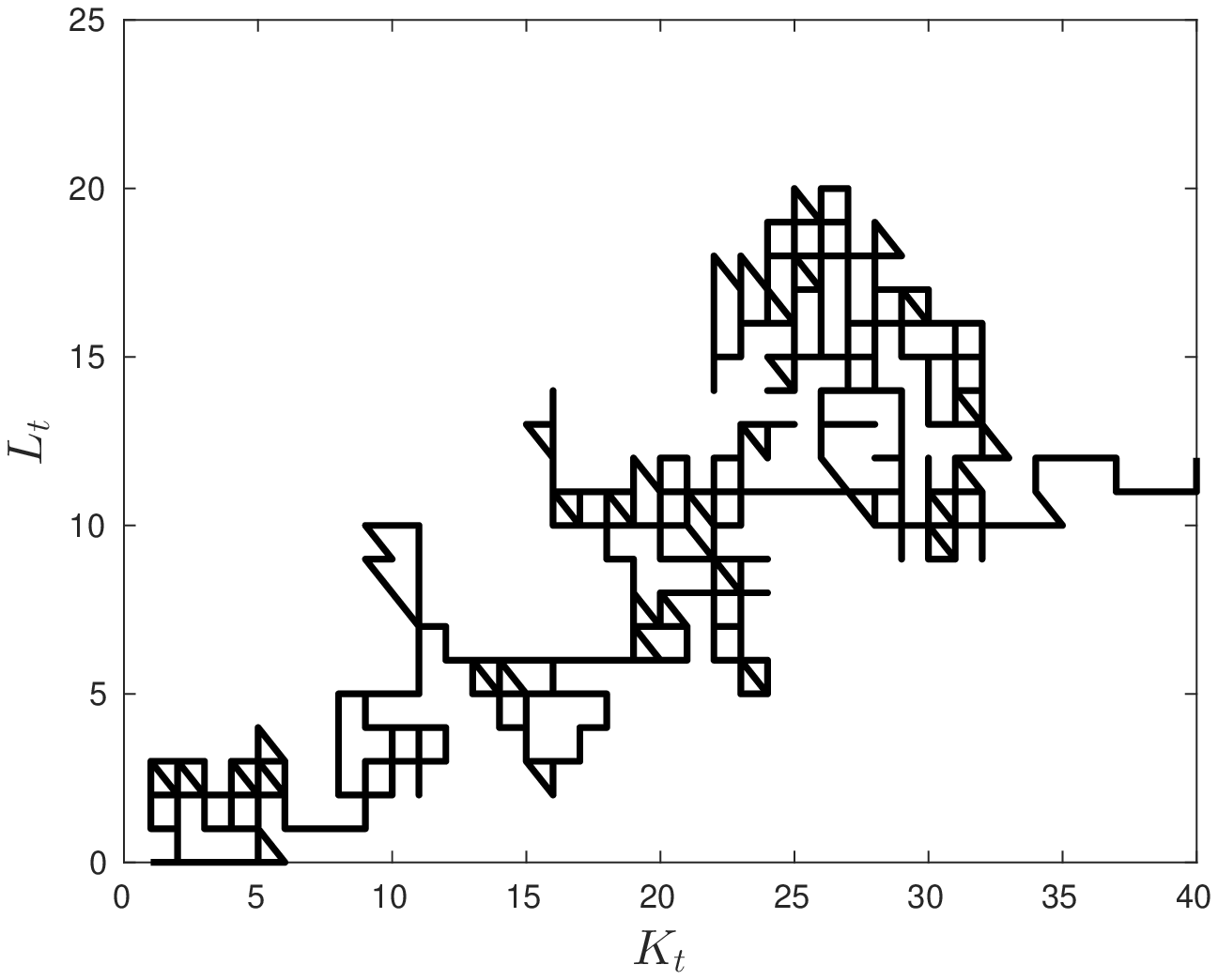}
\includegraphics[width=0.443\textwidth]{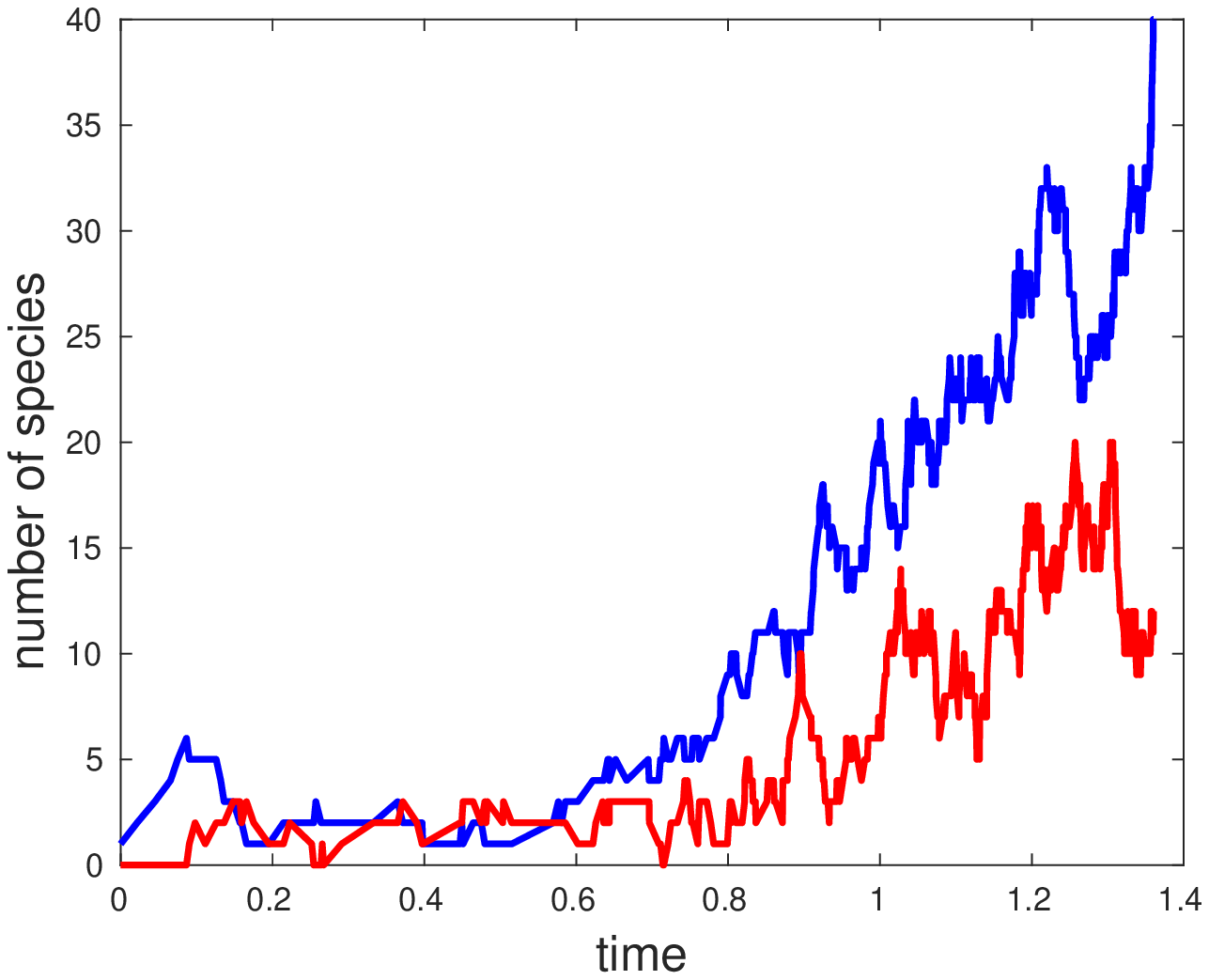}}
\caption{Simulation of a two-type branching process 
with $\omega_0=0.075$,  $\omega_1=0.353$, 
$\lambda_1=0$, $\lambda=10.462$, $ \mu=8.526$, 
  $ p=0.732$, $\delta=11.623$,  $\mu_0=5.411$ 
  and $\mu_1=25.468$. The left plot gives the trace of $(K_t,L_t)$ in the
$(k,\ell)$ plane; Plot on the right shows the paths over time of
$K_t$ and $L_t$ in blue and red, respectively. }
\label{fig:branchsim2}  
\end{figure*}

\noindent These are linked to the mutation rates 
\[
\omega_0=\frac{\mu_0}{c}=0.075, \quad \omega_1=\frac{\mu_1}{c}=0.353, \quad c=72.147,
\]
such that, at time $t=1$,
\[
dN/dS|_1= q\omega_0+(1-q)\omega_1=0.0955,\quad  q=0.9263.
\]
The corresponding two-type branching process is supercritical with
extinction probability $\mu/\lambda= 0.815$. Figure \ref{fig:branchsim2} shows a fairly typical
realization of the branching process, conditional on
non-extinction. Left panel of Figure \ref{fig:branchsim2} gives the trace of $(K_t,L_t)$ in the
$(k,\ell)$ plane, whereas the right panel shows the paths over time of
$K_t$ and $L_t$.

\subsection*{Solanaceae Family}

\begin{figure*}[!t]
\centerline{\includegraphics[width=0.99\textwidth]{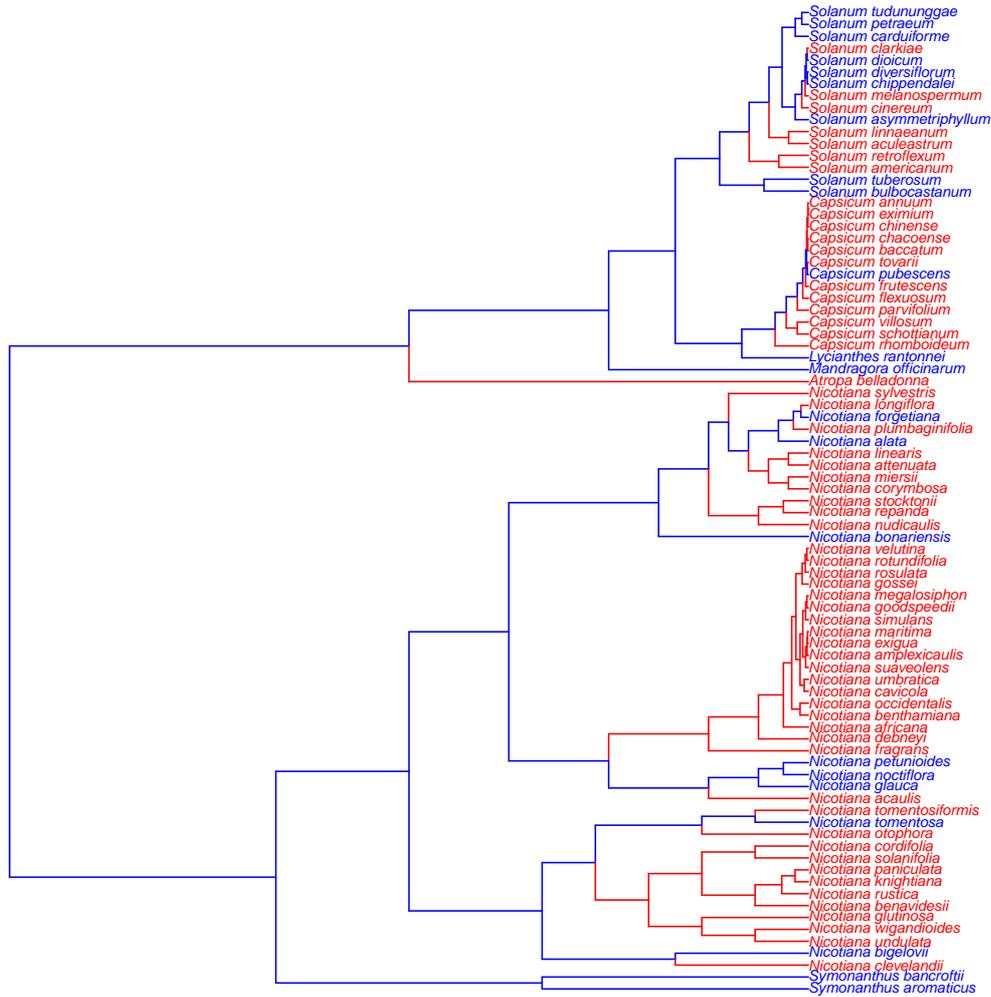}}
\caption{Ultrametric phylogenetic tree consisting of $83$ 
species in the Solanaceae family. Outcrosser 
branches are colored blue; the selfer branches, from bifurcation time onward,
are given in red.}
\label{soln_phylogeny_ultra}
\end{figure*}

The data set for the Solanaceae family, obtained from \cite{glemin_muyle}, consists 
of a total of $1527$ codons in an interleaved format and  
$m=83$ species ($k=22$ outcrossers and $\ell=61$ 
selfers). The selfing species form $25$ separate clusters. The 
corresponding phylogenetic tree is given in 
Figure \ref{soln_phylogeny_ultra}. 

It is straightforward to repeat all steps $1-3$ outlined in the
previous section.  The $dN/dS$-ratio of the complete
species tree at $t=1$, is observed to be $Q=0.439$, whereas the
estimated $dN/dS$-value restricted to outcrossers is
$\omega_0=0.4206$. The ultrametric version 
(Fig. \ref{soln_phylogeny_ultra}) of the phylogenetic tree, when the
`virtual' root length is set at $0.1$, has total branch length 
$T_\mathrm{tot}=7.881$. The sum of all outcrossing branch 
lengths (including the root) is
$T^{(0)}_\mathrm{tot}=4.845$. 
Hence, the minimal fraction of outcrossers 
is approximately $q^0= 0.615$ 
and therefore, as in (\ref{om1bound}),
\[
\omega_1\ge \frac{Q- \omega_0 q^0}{1-q^0}=0.468.
\]
The estimation of admissible pairs $(\lambda,\mu)$ proceeds by
applying the following methods.

\noindent
i) Use (\ref{outnumber}) with $k=22$ as
\begin{equation}\nonumber
\frac{\lambda e^{(\lambda-\mu)}-\mu}{\lambda-\mu}=22.
\end{equation}
ii) Use (\ref{bifurctime}) with $\bar{s}=0.183$, 
obtained from the bifurcation times $(s_1,\dots,s_{21})$ 
of the outcrossing species, that is
\begin{align*}
\int_0^1(1-F_1(s))\,ds \\=\int_0^1 \big(1-&\frac{\lambda-\mu
  e^{-\gamma_0}}{1-e^{-\gamma_0}}\,\frac{1-e^{-\gamma_0
    s}}{\lambda-\mu e^{-\gamma_0 s}}\Big)\,ds \nonumber = 0.183. 
\end{align*}
\noindent
iii) Use (\ref{branchestimation}) with $T^{(0)}_\mathrm{tot}=4.845$ as follows
\begin{equation}\nonumber
\int_0^1 \frac{\lambda e^{\gamma_0}-\mu}{\lambda
  e^{\gamma_0s}-\mu}\,ds=4.845. 
\end{equation}
The above three cases give an approximation for $\lambda$ 
and $\mu$ values, i.e.,  $\lambda\approx 8.097$ 
and $\mu\approx 6.417$.

From here on, attempting to estimate the remaining parameters
 $\omega_1$, $p$, $\delta$, and $\lambda_1$ (which will give $\mu_0$
and $\mu_1$), it becomes apparent that the Solanaceae data set does not line up
with the branching model as accurately as the Geraniaceae data set did.  Our tools
rely on measuring the number of selfers, the number of selfing clusters, and
the length of selfing branches.  Any two of these can be matched to a set of
parameters of the branching tree model, but matching all three
together seem to be out of range. For example, if we
focus on number of selfers and number of clusters, the 
fraction of selfing branch length, $1-q$, arising in the branching
model, will be larger than the fraction accounted for in the observed
ultrametric tree.  An interpretation of the discrepancy might be that
the sequence data set is so restricted that many of the selfers come
out biased towards short divergence times. These difficulties remain even if we
allow for more flexible conditions than (\ref{c}). As an example, the parameters
\[
\mu=13.3,\;  \lambda=14.1,\; \lambda_1=9.5,\;  p=0.25,\;  \delta=7.5,\;  \mu_0=7.675,
\]
are consistent with the $dN/dS$ value in selfers, that is $\omega_1=0.6$, as well as the
correct number of selfers and selfing clusters in the tree. 
The corresponding two-type branching process is supercritical with
extinction probability $\mu/\lambda= 0.943$.

\section{Discussion}

In this paper, we combined trait-dependent 
diversification models, BiSSE and ClaSSE, with 
trait-dependent substitution models. We focused 
on binary-trait models, based on the `evolutionary 
dead-end' hypothesis, with unidirectional shift 
from a trait-$0$ to a trait-$1$, where trait-$1$ 
had a lower diversification rate than trait-$0$. 
To do so, we first described several properties 
of the BiSSE and ClaSSE models that have not 
been obtained before. In particular, we introduced 
a novel way to decompose and analyze the reduced 
trait-$0$ and trait-$1$ species trees. We then showed how 
trait-dependent diversification processes affect the 
inference and interpretation of relationships between 
traits and molecular evolutionary rates.

We obtained several expressions describing the tree 
characteristics, such as the expected branch lengths 
of the reduced trees as well as their cluster sizes. 
The expected number of trait-$1$ species per cluster in particular, 
is biologically relevant; numerical exploration showed 
that this value stayed within controlled boundaries under a 
wide range of conditions. Assigning outcrossing and selfing 
species as trait-$0$ and trait-$1$, respectively, this formally 
confirms the general observation that large clades of selfing 
species are rare. Larger clusters of selfing species can only 
be obtained for the supercritical case, i.e., if $\lambda_1>\mu_1$, 
which corresponds to conditions where 
selfing is no more an evolutionary dead-end.

A specially interesting case occurs when selfing species 
are only found as singletons, as shown in the 
Geraniaceae family example. Such a case can be 
rather frequent -- for example it corresponds to $5$ of 
the $16$ datasets in \cite{glemin_muyle} study -- 
and it allows for a more complete treatment of the problem. 
In particular, we showed that under this condition, testing 
the difference of $dN/dS$ between outcrossing and selfing 
branches can be difficult. For longer branches leading to the 
selfing species, we could expect to have more power to 
distinguish between $\omega_0$ and $\omega_1$, simply 
because more substitutions occur and $\omega_0$, $\omega_1$ 
values are better estimated. However, the proportion of the branch 
with $\omega_1$ decreases as the branch length increases, 
making $\omega_0$ and $\omega_1$ values closer to one another. 
No matter what the conditions may be, assuming a pure cladogenetic model, 
when it is not, reduces the ability to detect differences in 
$\omega_0$ and $\omega_1$ values, either by lack of power 
(short branches), or, by incorrect trait assignation (long branches). 
Moreover, for short branches, the mutation substitution process 
cannot be considered as instantaneous, and the polymorphic 
transitory phase -- currently not included in our modeling framework --  
must be taken into account to avoid detection of spurious changes 
in $dN/dS$ \cite{mugal_etal}. Thus, not detecting any effect of 
selfing on $dN/dS$ does not necessarily mean that the effect is weak. 
The same rationale and conclusions can be applied to other binary traits, 
such as hermaphroditism versus dioecy \cite{kafer_etal}, sexuality 
versus clonality \cite{henry_etal} and solitary life versus sociality 
\cite{romiguier_etal}.

Overall, our results show that trait-dependent diversification processes 
can have a strong impact on the relationship between traits and molecular evolution. 
A further step would be to develop statistical methods allowing to jointly 
infer the effect of traits on both the diversification process and, the 
molecular evolutionary rates. Here, we propose a straightforward way to 
connect the two processes by making substitution and extinction rates 
proportional to one another, but this could be extended to other functions -- a linear 
function of course being a more natural assumption. Allowing 
for two-sided transitions (reversion from selfing to outcrossing, in our example) 
would also be a natural extension, but would lead to more complex 
treatments, since both trait-$0$ and trait-$1$ trees could be disjoint. 
We hope that the modeling framework presented in our paper, 
will be a useful starting point for further development in this field of research.

\section*{Supplementary Material}
The Maple code use to obtain Figures \ref{lambdamu} and \ref{fig:gerparest}
can be found in the Dryad Digital Repository: 
http://dx.doi.org/10.5061/dryad.3rg40

\section*{Funding}

This work was supported by the Marie Curie Intra-European Fellowship 
(grant number IEF-623486, project SELFADAPT to S.G. and M.L.). D.T. is 
supported by The Centre for Interdisciplinary Mathematics in 
Uppsala University.

\section*{Appendix 1}

This section elaborates the mathematical properties of 
the two-type, continuous time Markov branching process 
$X_t=(K_t,L_t)$, given in (\ref{branchingprocess}). 
Here, we analyze the process following the same approach 
and notation as in \cite{athreya_ney}.

The branching rates of $X_t$, given in (\ref{branchingrates}), are rewritten below for convenience 
\begin{equation}
(k,\ell)\mapsto \left\{
\begin{array}{cc}
(k+1,\ell) & \lambda_0 k\\
(k-1,\ell+1) & (1-p)\delta k\\
(k-1,\ell) & \mu_0 k\\
(k,\ell+1) & p\delta k+\lambda_1 \ell\\
(k,\ell-1) & \mu_1 \ell.
\end{array}
\right. \tag{\ref{branchingrates}}
\end{equation}
The life length of type-$i$, $i=0,1$, is exponentially distributed 
with parameter $\mathbf{a}=(a_0,a_1)$, such that
\[
a_0=\lambda_0+\mu_0+\delta\quad \mbox{and} \quad a_1=\lambda_1+\mu_1.
\]
The offspring distribution of the two types is given by 
$\mathbf{p(j)}=(p^{(0)}(\mathbf{j}),p^{(1)}(\mathbf{j}))$, where
\[
p^{(0)}(2,0)=\frac{\lambda_0}{\lambda_0+\delta+\mu_0},\quad
p^{(0)}(0,1)=\frac{(1-p)\delta}{\lambda_0+\delta+\mu_0},
\]
\[
p^{(0)}(0,0)=\frac{\mu_0}{\lambda_0+\delta+\mu_0},\quad
p^{(0)}(1,1)=\frac{p\delta}{\lambda_0+\delta+\mu_0},
\]
\[
p^{(1)}(0,2)= \frac{\lambda_1}{\lambda_1+\mu_1},\quad \quad
p^{(1)}(0,0)=\frac{\mu_1}{\lambda_1+\mu_1},
\]
and  $$\sum_{j} p^{(i)}(\mathbf{j})=1.$$
The generating function is of the form 
$\mathbf{f(s)}=(\mathbf{f}^{(0)}(\mathbf{s}),\mathbf{f}^{(1)}(\mathbf{s}))$,
where $\mathbf{f}^{(i)}(\mathbf{s})=\sum_{j} p^{(i)}(\mathbf{j})\mathbf{s}^j$, 
that is
\[
f^{(0)}(s_0,s_1)=\frac{\lambda_0s_0^2+p\delta s_0s_1+(1-p)\delta s_1+\mu_0}{\lambda_0+\delta+\mu_0},
\]
and
\[
f^{(1)}(s_0,s_1)= \frac{\lambda_1s_1^2+\mu_1}{\lambda_1+\mu_1}.
\]
The infinitesimal generating function is given by 
$\mathbf{u}^{i}(\mathbf{s})=a_i[\mathbf{f}^{(i)}(\mathbf{s})-s_i]$. Hence,
\begin{equation}
\begin{aligned}\nonumber
u^{(0)}(s_0,s_1) ={} \lambda_0s_0^2+p\delta s_0s_1+ 
       (1-p)\delta s_1 -(\lambda_0+\delta+\mu_0)s_0+\mu_0,
\end{aligned}
\end{equation}
and
\[
u^{(1)}(s_0,s_1)=\lambda_1s_1^2-(\lambda_1+\mu_1)s_1+\mu_1.
\]
The mean offspring matrix $A$ is defined as
\[
A=(a_{ij}),\quad \mbox{where} \quad a_{ij}=a_i\Big[\frac{\partial
  f^{(i)}(\mathbf{s})}{\partial s_j}\Big|_{\mathbf{s}=1}-\delta_{ij}\Big],
\]
and
\[
\delta(i,j)= \left\{
\begin{array}{cc}
1 \quad \;\;\; \mbox{if} \;\; \quad i=j\\
0 \quad \mbox{otherwise}. 
\end{array}
\right.
\]
Hence,
\[
A=\left(
\begin{array}{cc}
 \lambda_0-(1-p)\delta-\mu_0& \delta\\
0 & \lambda_1-\mu_1
\end{array}\right).
\]
The eigenvalues of $A$,
\[
\gamma_0=\lambda_0-(1-p)\delta-\mu_0 \quad \mbox{and} \quad \gamma_1=\lambda_1-\mu_1,
\]
are classified as
\[
\gamma_+=\max(\gamma_0,\gamma_1)\quad \left\{
\begin{array}{lc}
<0 & \mbox{subcritical}\\
=0 & \mbox{critical}\\
>0 & \mbox{supercritical}.
\end{array}\right.
\]
The extinction probabilities
\[
q_0=P(\mbox{extinction}|\mbox{starting with one outcrosser}),
\]
and
\[
q_1=P(\mbox{extinction}|\mbox{starting with one selfer}),
\]
are the solutions of the system
\[
q_0=f^{(0)}(q_0,q_1),\quad q_1=f^{(1)}(q_0,q_1),
\]
or equivalently
\[
u^{(0)}(q_0,q_1)=0,\quad u^{(1)}(q_0,q_1)=0.
\]
By solving this system, we obtain
\[
q_0=\frac{\mu_0+(1-p)\delta}{\lambda_0} \quad \mbox{and} \quad  q_1=1.
\]

\paragraph{\textit{Mean values}.---} 
Given 

\[
m_{00}(t)=E(K_t|K_0=1), \quad \;\;\; m_{11}(t)=E(L_t|L_0=1),
\]
\[
m_{01}(t)=E(L_t|K_0=1), \;\;\; m_{10}(t)=E(K_t|L_0=1)=0,
\]

\[
M(t)=\left(
\begin{array}{cc}
m_{00}(t) & m_{01}(t)\\
0 & m_{11}(t)
\end{array}\right),
\]
we have $M(s+t)=M(s)M(t)$ and $M(t)\to I$ as $t\to 0$. Hence
$M(t)=e^{At}$, and
\[
m_{00}(t)=e^{(r_0-(1-p)\delta)t},\quad m_{11}(t)=e^{r_1t},
\]
\[
m_{01}(t)=\frac{\delta}{r_0-(1-p)\delta-r_1}\big[e^{(r_0-(1-p)\delta)t}-e^{r_1t}\big].
\]
Here, $M(t)$ is {\it not} positively regular.  However,
\[
\frac{E(K_t)}{E(K_t)+E(L_t)}=\frac{m_{00}(t)}{m_{00}(t)+m_{01}(t)}
\to 1-\frac{\delta}{r_0-r_1+p\delta},
\]
and
\[
\frac{E(L_t)}{E(K_t)}= \frac{m_{01}(t)}{m_{00}(t)} \to
\frac{\delta}{r_0-(1-p)\delta-r_1},
\]
as $t\to\infty$.

\end{document}